\documentclass[12pt]{elsarticle}
\usepackage{amssymb}
\usepackage{amsmath}%[fleqn]
\usepackage{amsthm}
\usepackage{subfigure}
\usepackage{graphicx}

\biboptions{numbers,sort&compress}

\usepackage[dvipdfm,colorlinks,linkcolor=red,anchorcolor=blue,citecolor=blue]{hyperref}
\usepackage{booktabs}
\makeatletter

\newcommand{\Rmnum}[1]{\expandafter\@slowromancap\romannumeral #1@}
\makeatother
\usepackage{caption}

\usepackage{anysize}
\usepackage{setspace}
\marginsize{1.5cm}{1.5cm}{0.6cm}{0.6cm} %%LRTB
\journal{XXX}

\begin{document}

\begin{frontmatter}

\title{Optical rogue waves and W-shaped solitons in the multiple self-induced transparency system }

\author{Xin Wang$^{{\rm a*}}$}\ead{wangxinlinzhou@163.com}
\author{Chong Liu$^{{\rm b}}$}
\cortext[cor1]{Corresponding author.}

\address{$^{{\rm a}}$Department of Mathematics, Zhongyuan University of Technology, Zhengzhou, 450007, China}
\address{$^{{\rm b}}$School of Physics, Northwest University, Xi\rq an, 710069, China}

\begin{abstract}
We study localized nonlinear waves on a plane wave background in the multiple self-induced transparency (SIT) system,
which describes an important enhancement of the amplification and control of optical waves
compared to the single
SIT system.
A hierarchy of exact multiparametric rational solutions in a compact determinant representation are presented.
We demonstrate that, this family of solutions contains known rogue wave solution and unusual W-shaped soliton solution, which strictly corresponds to the
linear stability analysis that involves modulation instability and stability regimes in the low perturbation frequency region.
State transitions between rogue waves and W-shaped solitons as well as the higher-order nonlinear superposition modes are revealed by the suitable choice for the background wavenumber of electric field component.
In particular, our results show that, the multiple SIT system admits stationary and nonstationary nonlinear modes in contrast to the results in the single
SIT system. Correspondingly, the important characteristics of the nonlinear waves including trajectories and spectrum are revealed in detail.
\end{abstract}


\end{frontmatter}

\section{Introduction }  %%% 节标题 2

Fiber optic communication has been a topic of considerable interest with important applications in
theoretical and experimental physics \cite{01}. Dissipation and dispersion effects are the main
hindrance for the signal propagation through optical fibers  \cite{02}.
To deal with this, one type of optical soliton that
is described by the nonlinear Schr\"{o}dinger (NLS) equation, where the group velocity dispersion in the light pulse wave guide
can be balanced by the self-phase modulation, has been extensively investigated \cite{03}.
The other type of coherent optical soliton that is based on self-induced transparency (SIT) firstly discovered by McCall and Hahn
in a pure two-level atomic transition system, has also received long-running attention \cite{04}.

It is discovered by Lamb that the SIT system (or Maxwell-Bloch equations) \cite{05}
\begin{subequations}\label{01}
\begin{align}
&E_{z}=<p>,\\
&p_{t}=EN+2i\omega p,\\
&N_{t}=-2(Ep^{*}+E^{*}p).
\end{align}
\end{subequations}
serves as a basic model to govern the SIT phenomenon in a two-level medium. Here, $E$ is the slowly varying complex envelope of
electric field, $<p>$ denotes the average polarization
$$
<p>=\int^{\infty}_{-\infty}p(t,z;\omega)g(\omega)d\omega,\ \int^{\infty}_{-\infty}g(\omega)d\omega=1,
$$
where $g(\omega)$ is the normalized probability density, $N$ stands for the population inversion between two level,
and the index $*$ represents complex conjugation.
The above system can be immediately converted to the AB system in geophysical fluid dynamics \cite{06} and
sine-Gordon equation in differential geometry \cite{07}.
Thus far, there has been a surge of great many reports on the
integrability and analytic solution of system (\ref{01}). For example,
the inverse scattering theory for solving the
the general initial value problem was established by Ablowitz and Gabitov et al. \cite{08,09}, and the Darboux transformation (DT)
for finding the soliton solution, exulton-like solution and  rational solution was constructed by Matveev and
Salle \cite{10}. In particular, more recently, rogue wave solution of system (\ref{01}) or AB system has been
widely researched  \cite{11,12,13,14}.

Remarkably, a significant enhancement of the amplification and control of the optical soliton in system (\ref{01}) can be
achieved if the single SIT system is generalized to the multiple case.  In this connection,
Kundu considered a multiple SIT system \cite{15}:
\begin{subequations}\label{02}
\begin{align}
&E_{z}=<p>,\label{2a}\\
&p_{t}=-e+EN+2i\omega p,\label{2b}\\
&N_{t}=-2(Ep^{*}+E^{*}p),\label{2c}\\
&e_{t}=-2EM+2i\omega e,\label{2d}\\
&M_{t}=-Ee^{*}+E^{*}e.\label{2e}
\end{align}
\end{subequations}
Here, $e$ and $M$ are the new introduced induced polarization and population inversion, respectively.
When taking $e\rightarrow0$ and $M\rightarrow0$, system (\ref{02}) is reduced to the single SIT system (\ref{01}).
The complete integrability and soliton solution of system (\ref{02}) have been discussed in Kundu\rq~work.

Nevertheless, to our knowledge, optical rogue waves and W-shaped solitons in system (\ref{02})
have not been reported anywhere.
As is known to all, rogue waves are defined as waves having large amplitudes and emerging
randomly in the particular dynamic system with low probability \cite{16}. The rogue wave phenomena were
originally applied to describe the unexpected appearance of huge and short-lived catastrophic waves on the ocean, but now
have rapidly spread to a wide range of physical systems including optical fibres  \cite{17},
Bose-Einstein condensates  \cite{18}, plasmas \cite{19}, cold atoms \cite{20} and so on.

The simplest simulation of a single rogue wave in mathematics is the Peregrine soliton, which is a rational solution
describing an \lq\lq amplitude peak\rq\rq~isolated in both space and time of the NLS equation \cite{21}.
Subsequently, it is demonstrated that the intricate and diverse types of higher-order rogue waves
which are constructed as the nonlinear superposition or combination of the first-order Peregrine rogue waves can exist in the
NLS equation  \cite{22,23}. However, it is necessary to say that, in order to model the complicated rogue wave phenomena in various
physical systems in a relevant way, one should transcend the standard NLS description.
Consequently, some important NLS-type equations (e.g. Hirota equation and Sasa-Satsuma equation) with
higher-order and/or dissipative effects  \cite{24,25}, and the coupled-wave systems (e.g. Manakov system and coupled Hirota equations)
with interacting wave components of different amplitudes and frequencies  \cite{26,27,28} have been of continued interest.
Most recently, Akhmediev and one of the authors Liu show that the higher-order integrable perturbations
and coupled-wave interaction can indeed generate state transition effects from breather/rogue wave to
W-shaped soliton  \cite{29,30,31,Ren}.

The purpose of this paper is mainly concentrated on two aspects: (1) we construct an $N$th-order rational solution
in a compact determinant representation by taking advantage of the generalized DT approach \cite{32,33,34,35};
(2) With the aid of the analytical rational solution and modulation instability (MI),
dynamics of the optical rogue waves, as well as the stationary and nonstationary
W-shaped solitons from first to third order are presented. In particular,
it is importantly found that, the nonstationary W-shaped solitons can only exist in the multiple SIT system,
while for the single system they are impossible to appear.

Our paper is organized as follows: section 2 derives a compact determinant form for an $N$th-order rational solution
by using the generalized DT \cite{32,33,34,35}. In section 3, dynamics of different kinds of rogue waves and W-shaped solitons,
and the important state transition is revealed. Moreover, the trajectories of the rogue waves are shown.
In section 4, spectrum and energy of the rogue wave solution are given.
At last, the conclusion and some discussion are given.

\section{$N$th-order rational solution }
Our starting point is the Lax pair of system (\ref{02}), which can be constructed in the form
\begin{align}
&\Psi_{t}=U\Psi,\label{03}\\
&\Psi_{z}=V\Psi,\label{04}
\end{align}
where
$$
\begin{array}{l}
U=i\left(
    \begin{array}{cc}
      \zeta & 0 \\
      0 & -\zeta \\
    \end{array}
  \right)+\left(
    \begin{array}{cc}
      0 & E \\
      -E^{*} & 0 \\
    \end{array}
  \right),\\
V=-\dfrac{1}{4(\zeta-\omega)^2}\left(
                                \begin{array}{cc}
                                  M & e \\
                                  -e^{*} & -M \\
                                \end{array}
                              \right)+
\dfrac{1}{4i(\zeta-\omega)}\left(
                            \begin{array}{cc}
                              N & -2p \\
                              -2p^{*} & -N \\
                            \end{array}
                          \right).
\end{array}
$$
Here, $\zeta$ is the complex eigenvalue parameter. It is readily to prove that system (\ref{02}) can be exactly reproduced
from the compatibility condition $U_{z}-V_{t}+UV-VU=0$.

To proceed, let $\Psi_{1}=(\psi_{1},\phi_{1})^{T}$ be a basic solution of Eqs. (\ref{03}) and (\ref{04}) with
$E=E[0]$, $p=p[0]$, $N=N[0]$, $e=e[0]$, $M=M[0]$ and $\zeta=\zeta_{1}$, then one can verify that the following
gauge transformation
\begin{align}
&\Psi[1]=T\Psi,\ T=I+\frac{\zeta_{1}^{*}-\zeta_{1}}{\zeta-\zeta_{1}^{*}}\frac{\Psi_{1}\Psi_{1}^{\dag}}{\Psi_{1}^{\dag}\Psi_{1}},\label{05}\\
&E[1]=E-2i(\zeta_{1}^{*}-\zeta_{1})\frac{\psi_{1}\phi_{1}^{*}}{|\psi_{1}|^2+|\phi_{1}|^2},\label{06}
\end{align}
is just the classical DT of the Lax pair system.  Here, $\dag$ denotes Hermite conjugation.
We point out that, in this paper, we only pay our attention to the rational solutions
including rogue wave solution and W-shaped soliton solution
of the electric field, i.e. the $E$ component. For the $p$, $N$, $e$ and $M$ components, we omit
the results, since their new potentials can be directly derived from the differential
and/or integral calculations of $E[1]$ by virtue of the original system, and
the corresponding rogue waves and W-shaped solitons in these components are similar to the ones in $E$
component with the same initial excitations.

After that, we introduce a general plane-wave solution of system (\ref{02}), that is
\begin{equation}\label{07}
\begin{array}{l}
E[0]=ce^{i\theta},\ p[0]=icbe^{i\theta},\ \theta=at+bz, \ N=d_{0},\\
e[0]=c(ab-2b\omega+d_{0})e^{i\theta},\ M[0]=-1/2i(ab-2b\omega+d_{0})(a-2\omega).
\end{array}
\end{equation}
Here, $c$, $a$ and $b$ represent the background amplitude, frequency and wave number of the electric field, respectively, and
$d_{0}$ is a real constant.

Next, in view of the classical DT (\ref{05}), (\ref{06}) and its higher-degree generalized form
by taking the limit technique,
we can directly work out a unified $N$th-order rational solution of
the following compact form as
\begin{equation}\label{solu}
E[N]=\left(1-2i\frac{\det(H_{1})}{\det(H)}\right)ce^{i\theta},
\end{equation}
where
\begin{align}
&H=\left(
     \begin{array}{cccc}
       H[1,1] & H[1,2] & \cdots & H[1,N] \\
       H[2,1] & H[2,2] & \cdots & H[2,N] \\
       \vdots & \vdots & \ddots & \vdots \\
       H[N,1] & H[N,2] & \cdots & H[N,N] \\
     \end{array}
   \right),\nonumber\\
&H_{1}=\left(
     \begin{array}{ccccc}
       H[1,1] & H[1,2] & \cdots & H[1,N] & \phi_{1}^{[0]*}\\
       H[2,1] & H[2,2] & \cdots & H[2,N] & \phi_{1}^{[1]*}\\
       \vdots & \vdots & \ddots & \vdots & \vdots\\
       H[N,1] & H[N,2] & \cdots & H[N,N] & \phi_{1}^{[N-1]*}\\
       \psi_{1}^{[0]}/c & \psi_{1}^{[1]}/c &\cdots &\psi_{1}^{[N-1]}/c &0
     \end{array}
   \right),\nonumber
\end{align}
with
\begin{align}
&H[i,j]=\dfrac{1}{2(i-1)!2(j-1)!}\dfrac{\partial ^{2(i+j-2)}}{\partial f^{2(j-1)}\partial f^{*2(i-1)}}
\dfrac{\psi_{1}\psi_{1}^{*}+\phi_{1}\phi_{1}^{*}}{2ic(1+f^2+f^{*2})}\bigg|_{f,f^{*}\rightarrow0},\nonumber\\
&\psi_{1}^{[l]}=\dfrac{1}{2l!}\dfrac{\partial^{2l}\psi_{1}}{\partial f^{2l}}\bigg|_{f\rightarrow0},\
\phi_{1}^{[l]}=\dfrac{1}{2l!}\dfrac{\partial^{2l}\phi_{1}}{\partial f^{2l}}\bigg|_{f\rightarrow0},\ l=0,1,\cdots,N-1,\nonumber
\end{align}
being denoted by
\begin{align}
&\psi_{1}=(C_{2}e^{A}-C_{1}e^{-A})e^{\frac{i}{2}\theta},\nonumber\\
&\phi_{1}=(C_{1}e^{A}-C_{2}e^{-A})e^{-\frac{i}{2}\theta},\nonumber
\end{align}
such that the coefficients satisfy
$$
C_{1}=\frac{(1+2f^2+2f\sqrt{1+f^2})^{\frac{1}{2}}}{2f\sqrt{1+f^2}},\
C_{2}=\frac{(1+2f^2-2f\sqrt{1+f^2})^{\frac{1}{2}}}{2f\sqrt{1+f^2}},
$$
and
$$
A=2cf\sqrt{1+f^2}\bigg[t-\frac{\bigg(ab-4b\omega+d_{0}+2b(\dfrac{a}{2}+ic+2icf^2)\bigg)}{4\bigg(\dfrac{a}{2}
+ic+2icf^2-\omega\bigg)^2}z+\sum_{i=1}^{N-1}s_{i}f^{2i}\bigg].
$$
Here, $f$ is a complex small parameter, and $s_{i}$ are the complex constants.

\section{MI, rogue wave and W-shaped soliton }

\subsection{Modulation instability}
It has been recently realized that rogue wave exists only in the MI subregion with zero-frequency  \cite{36}, and the
state transition between rogue wave and W-shaped soliton arises from the attenuation of MI growth rate \cite{30,37}.
Hence for the sake of using Eq. (\ref{solu}) to describe the dynamics of rogue wave and W-shaped soliton,
we first perform the standard MI analysis to reveal the MI characteristic arising from the coupling effects.

To this end, we consider the following equivalent form of system (\ref{02})
\begin{subequations}\label{09}
\begin{align}
&E_{ttz}=4i\omega E_{tz}+(EN)_{t}+4\omega^2 E_{z}+2EM-2i\omega EN,\\
&N_{t}=-2(|E|^2)_{z},\\
&M_{t}=EE^{*}_{tz}-E^{*}E_{tz}+2i\omega (|E|^2)_{z},
\end{align}
\end{subequations}
where the $p$ and $e$ components are eliminated by substituting Eqs. (\ref{2a}) and (\ref{2b}) into Eq. (\ref{2d}).

At this point, a perturbed nonlinear background may be given by
$$
E=ce^{i\theta}(1+u),\ N=d_{0}+v,\ M=-1/2i(ab-2b\omega+d_{0})(a-2\omega)+iw,
$$
where $u(t,z)$, $v(t,z)$ and $w(t,z)$ are the small perturbed functions and fulfil a linear equation group.
Let
$$
u=f_{+}e^{i\kappa(t-\Omega z)}+f_{-}^{*}e^{-i\kappa(t-\Omega^{*} z)},\
v=ge^{i\kappa(t-\Omega z)}+g^{*}e^{-i\kappa(t-\Omega^{*} z)},\
w=he^{i\kappa(t-\Omega z)}+h^{*}e^{-i\kappa(t-\Omega^{*} z)},
$$
where $\kappa$ and $\Omega$ are assumed to be real and complex.
Embedding them
into system (\ref{09}) and taking the linear part for $\alpha=(f_{+},f_{-},g,h)^{T}$, we have the algebraic equation
$B\alpha=0$, where
\begin{align}
&B_{11}=-\Omega\kappa^3+(-2a\Omega+4\Omega\omega+b)\kappa^2+(-a^2\Omega+4a\Omega\omega-4\Omega\omega^2+2ab-4b\omega+d_{0})\kappa,\nonumber\\
&B_{22}=\Omega\kappa^3+(-2a\Omega+4\Omega\omega+b)\kappa^2+(a^2\Omega-4a\Omega\omega+4\Omega\omega^2-2ab+4b\omega-d_{0})\kappa,\nonumber\\
&B_{31}=c^2\Omega\kappa^2+(ac^2\Omega-2c^2\Omega\omega-bc^2)\kappa,~ B_{32}=-c^2\Omega\kappa^2+(ac^2\Omega-2c^2\Omega\omega-bc^2)\kappa,\nonumber\\
&B_{12}=B_{21}=B_{33}=B_{44}=0,~B_{14}=B_{24}=2,~B_{41}=B_{42}=-2ic^2\Omega\kappa,\nonumber\\
&B_{34}=-\kappa, ~B_{43}=i\kappa,~B_{13}=\kappa+a-2\omega, ~B_{23}=-\kappa+a-2\omega. \nonumber
\end{align}

Solving $\det(B)=0$ we get the MI growth rate
\begin{equation}
G=|{\rm Im}\{\Omega\}|=\frac{\sqrt{4c^2-\kappa^2}}{(\kappa^2-a^2+4 a\omega-4 c^2-4\omega^2)^2}
\left|(\kappa^2-3 a^2+12a\omega-4 c^2-12\omega^2)b-2(a-2\omega)d_{0}\right|.
\end{equation}
Thus, it is easy to find that the MI exists in the region $-2c<\kappa<2c$. Particularly,
the modulation stability (MS) regions can be acquired analytically
\begin{equation}
b_{s}=\dfrac{2(a-2\omega)d_{0}}{\kappa^2-3 a^2+12a\omega-4 c^2-12\omega^2 }.
\end{equation}
The above MS representation depends on the background amplitude $c$,  the frequency $a$,
the constants $d_{0}$ and $\omega$, and the perturbation frequency $\kappa$.
In the following,
we will show that once the MI growth rate attenuates to zero in the zero-frequency MS
region, i.e. $b=b_{s}|_{\kappa=0}$, a transition between the rogue wave and W-shaped soliton occur.

\subsection{Rogue wave }

By utilizing Eq. (\ref{solu}) with $N=1$, we obtain the first-order rational solution which has the form
\begin{equation}\label{R1}
E[1]=ce^{i\theta}\dfrac{F_{1}+iG_{1}}{D_{1}},
\end{equation}
where
\begin{align}
&F_{1}=-4(a^2-4a\omega+4c^2+4\omega^2)^2c^2t^2+8(2a^3b-12a^2b\omega+24ab\omega^2-16b\omega^3+a^2d_{0}-4ad_{0}\omega\nonumber\\
&~~~-4c^2d_{0}+4d_{0}\omega^2)c^2zt-4(4a^2b^2-16ab^2\omega+4b^2c^2+16b^2\omega^2+4abd_{0}-8bd_{0}\omega+d_{0}^2)c^2z^2\nonumber\\
&~~~+3(a^2-4a\omega+4c^2+4\omega^2)^2,\nonumber\\
&G_{1}=-16(3a^2b-12ab\omega+4bc^2+12b\omega^2+2ad_{0}-4d_{0}\omega)c^2z,\nonumber\\
&D_{1}=4(a^2-4a\omega+4c^2+4\omega^2)^2c^2t^2-8(2a^3b-12a^2b\omega+24ab\omega^2-16b\omega^3+a^2d_{0}-4ad_{0}\omega\nonumber\\
&~~~-4c^2d_{0}+4d_{0}\omega^2)c^2zt+4(4a^2b^2-16ab^2\omega+4b^2c^2+16b^2\omega^2+4abd_{0}-8bd_{0}\omega+d_{0}^2)c^2z^2\nonumber\\
&~~~+(a^2-4a\omega+4c^2+4\omega^2)^2.\nonumber
\end{align}
Here we remark that the above rational solution contains two types of localized waves, i.e., (i)
rogue wave with $b\neq b_{s}|_{\kappa=0}$ and (ii) W-shaped soliton with  $b=b_{s}|_{\kappa=0}$.
Remarkably, this family of solutions contains known rogue wave solution and unusual W-shaped soliton solution, which strictly corresponds to the
above linear stability analysis that involves modulation instability and stability regimes in the low perturbation frequency region.

We first show the dynamics of the rogue wave.
In the case of $b\neq b_{s}|_{\kappa=0}$, Eq. (\ref{R1}) corresponds to the
celebrated \lq\lq eye\rq\rq-shaped (one hump and two valleys near $z=0$) rogue wave.
By using the trace analysis method,
we can define the motion of the rogue wave\rq~hump precisely
\begin{equation}\label{th}
T_{h}=Kz,
\end{equation}
where
$$
K=\dfrac{2(a-2\omega)^3b+(a-2c-2\omega)(a+2c-2\omega)d_{0}}{(a^2-4a\omega+4c^2+4\omega^2)^2},
$$
and the motions of its two valleys
\begin{equation}\label{tv}
T_{v}=\dfrac{K_{1}\pm 2\sqrt{K_{2}}}{4c(a^2-4a\omega+4c^2+4\omega^2)^2},
\end{equation}
where
$$
\begin{array}{l}
K_{1}=4(2a^3b-12a^2b\omega+24ab\omega^2-16b\omega^3+a^2d_{0}-4ad_{0}\omega-4c^2d_{0}+4d_{0}\omega^2)cz,\\
K_{2}=48(3a^2b-12ab\omega+4bc^2+12b\omega^2+2ad_{0}-4d_{0}\omega)^2c^4z^2+3(a^2-4a\omega+4c^2+4\omega^2)^4.
\end{array}
$$

\begin{figure}[!h]
\centering
\renewcommand{\figurename}{{\bf Fig.}}
{\includegraphics[height=6cm,width=8.5cm]{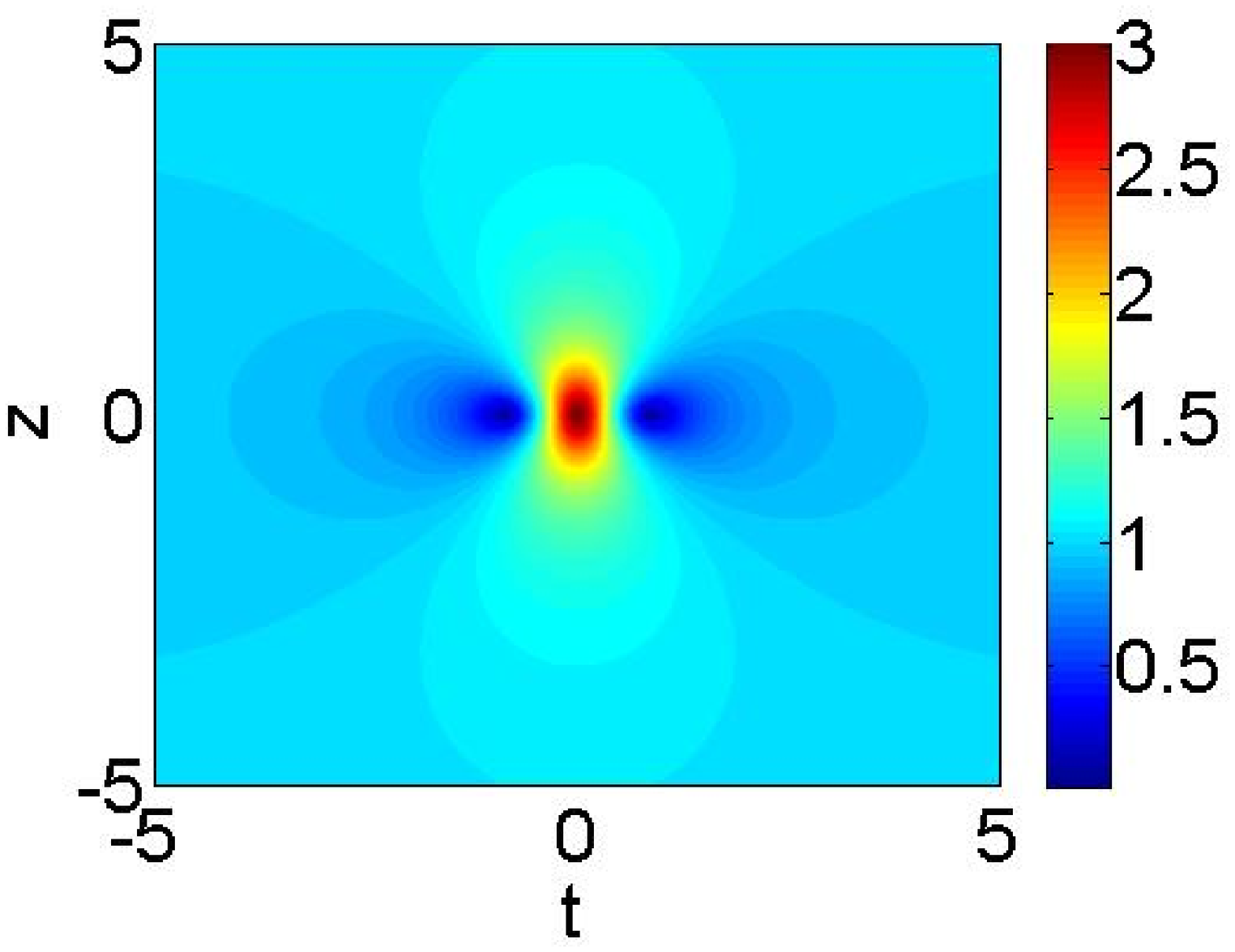}}
{\includegraphics[height=6cm,width=8.5cm]{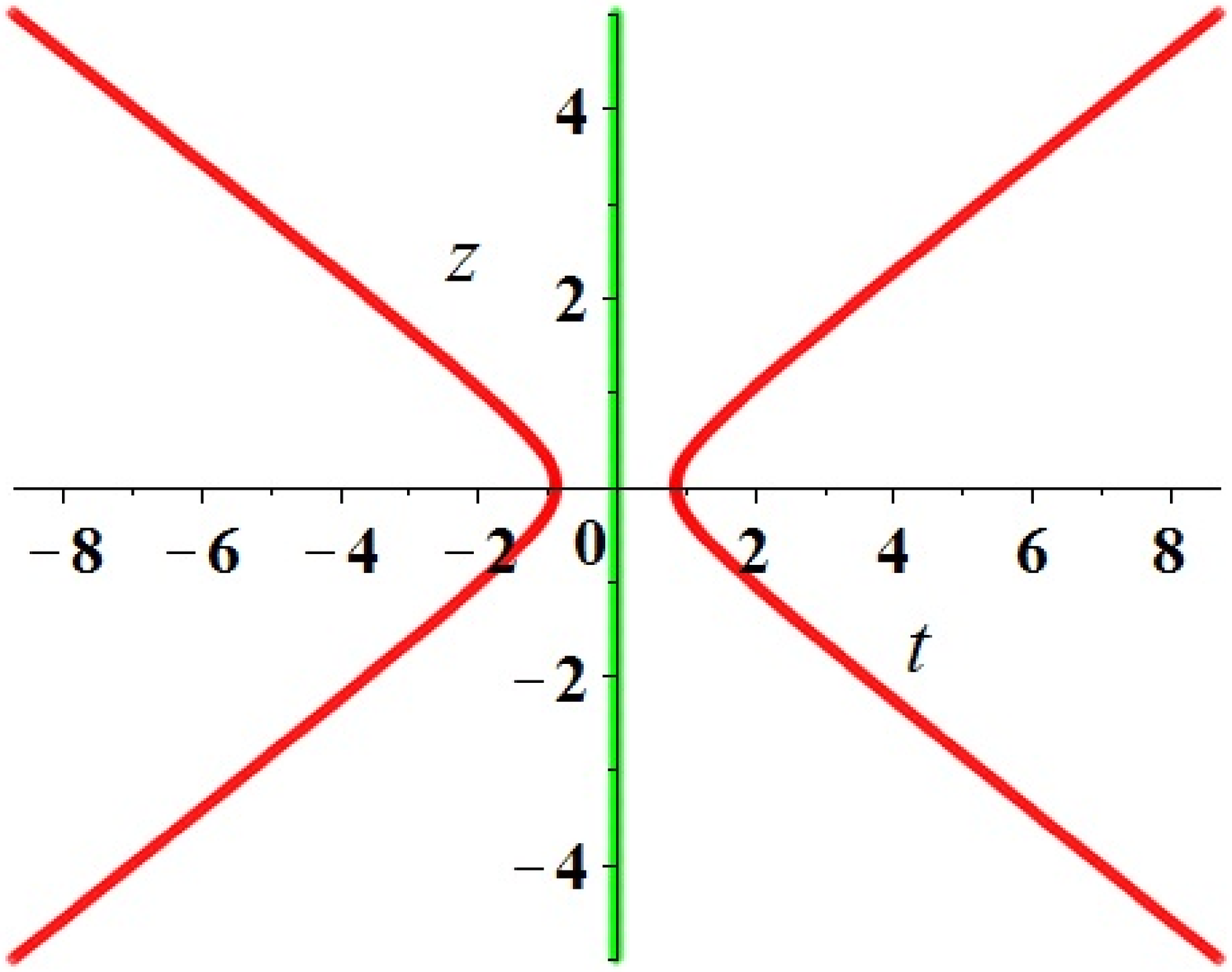}}
\begin{center}
\hskip 1cm $(\rm{a})$ \hskip 7cm $(\rm{b})$
\end{center}
\caption{(a) Density plot and (b) trace of the first-order rogue wave given by Eq. (\ref{R1}) with
$a=1,b=3/2,c=1,d_{0}=1,\omega=0$. The green and red
line correspond to the trajectories of the rogue wave\rq~hump and valleys, respectively. This holds for the other
pictures in this paper.}
\label{fig:1}
\end{figure}

Solving $K=0$ leads to
$$b=-\dfrac{(a-2c-2\omega)(a+2c-2\omega)d_{0}}{2(a-2\omega)^3}.$$
In this situation, we exhibit that in Fig. \ref{fig:1}(a), there are one hump and two valleys
in the temporal-spacial plane. The maximum amplitude of the rogue wave is 3 and is localized at $(0,0)$,
and the minimum amplitude of it is 0 and arrives at $(\pm\sqrt{3}/2,0)$.
Moreover, it is obvious to see that in Fig.  \ref{fig:1}(b),
the \lq\lq ridge\rq\rq~of the rogue wave is vertical to the $t$ axis, and the
trajectories of the two valleys look like an \lq\lq X\rq\rq~shape.
The rogue wave at this point is nothing but the standard Peregrine rogue wave
in the NLS equation \cite{34}.

\begin{figure}[!h]
\centering
\renewcommand{\figurename}{{\bf Fig.}}
{\includegraphics[height=6cm,width=8.5cm]{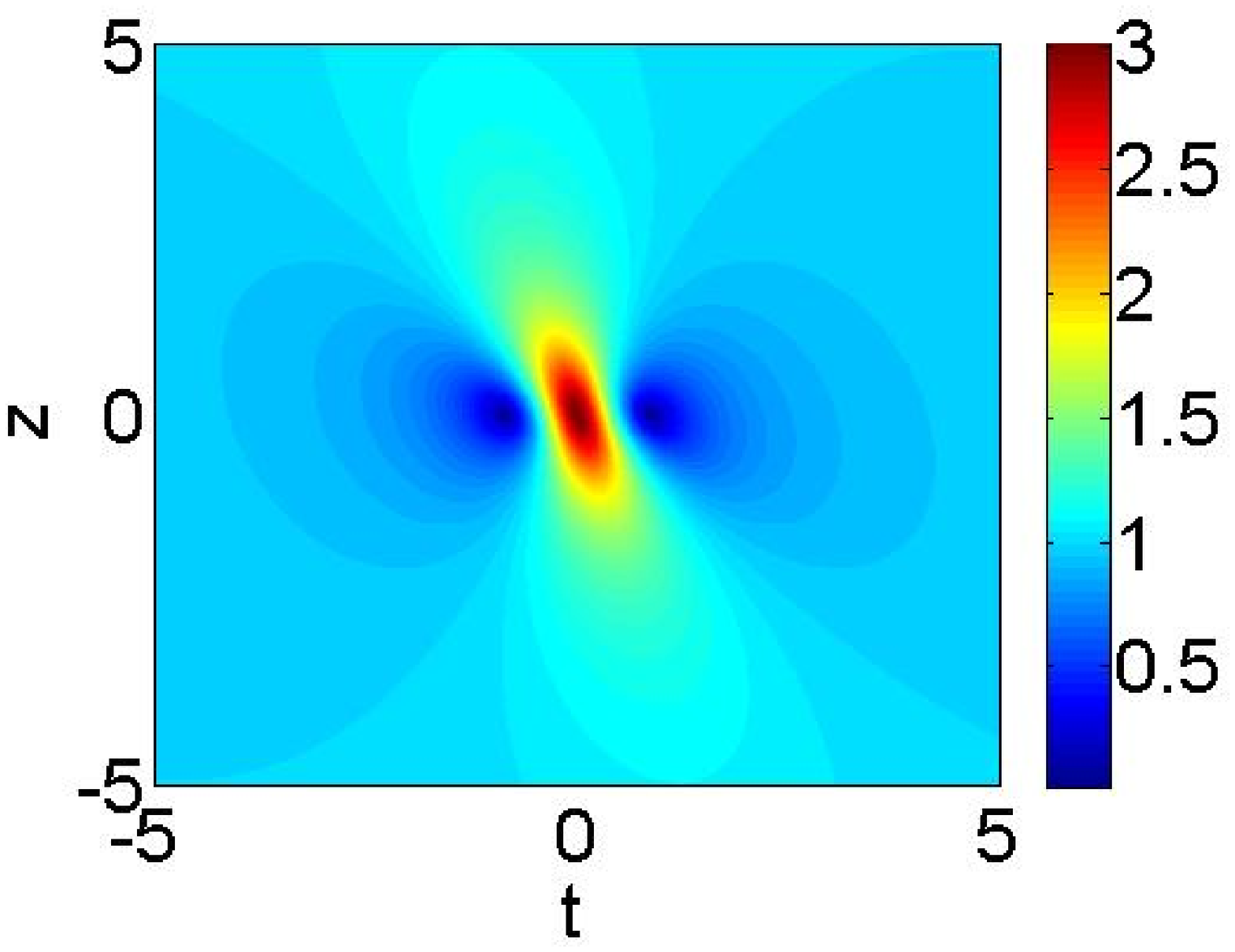}}
{\includegraphics[height=6cm,width=8.5cm]{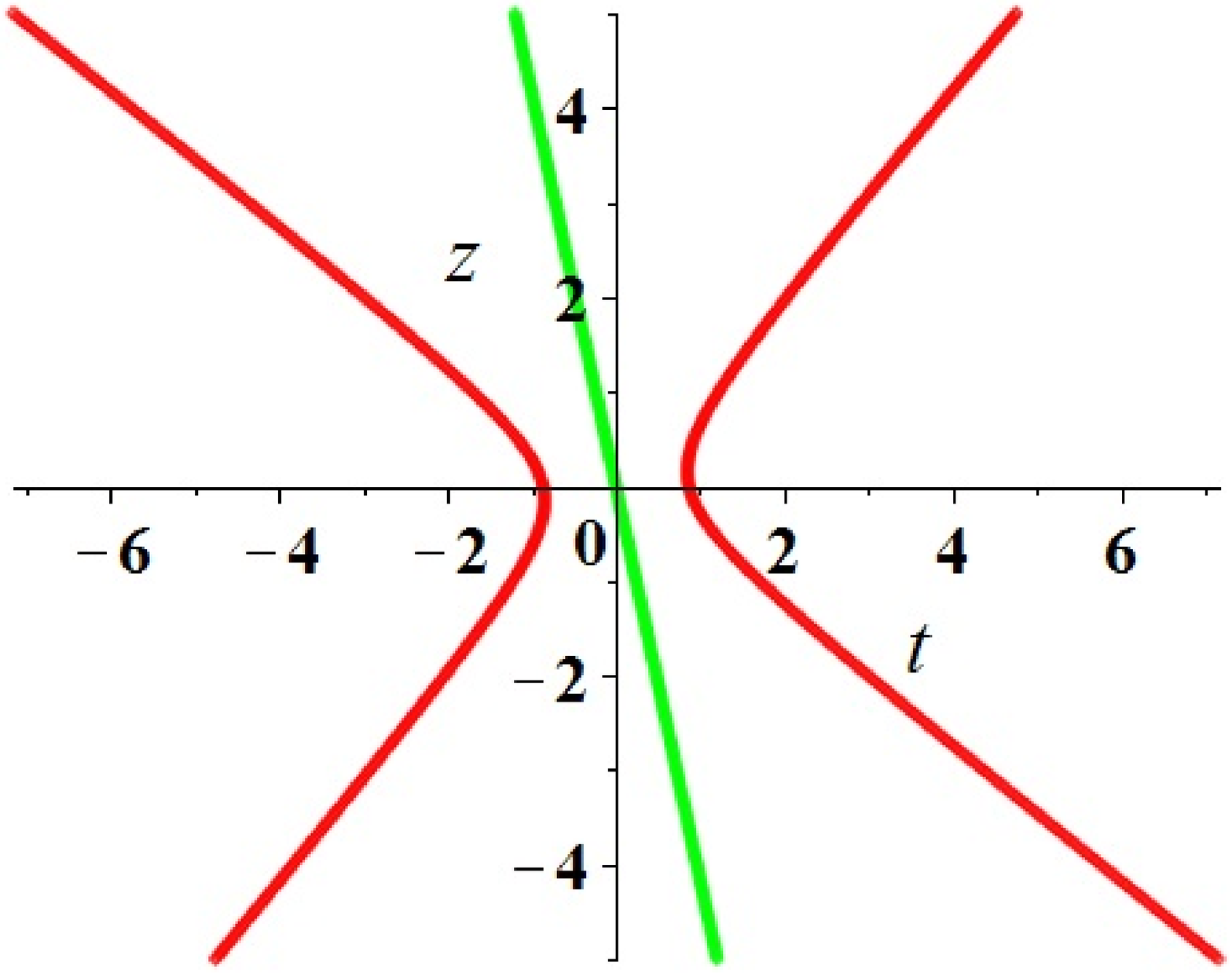}}
\begin{center}
\hskip 1cm $(\rm{a})$ \hskip 7cm $(\rm{b})$
\end{center}
\caption{(a) Density plot and (b) trace of the first-order rogue wave given by Eq. (\ref{R1})
with $a=1,b=-3/2,c=1,d_{0}=1,\omega=0$. }
\label{fig:2}
\end{figure}

When setting
$$b\neq-\dfrac{(a-2c-2\omega)(a+2c-2\omega)d_{0}}{2(a-2\omega)^3}$$
such that $K$ is not equal to 0. It is displayed that in Fig. \ref{fig:2}(a),
the basic shape of the rogue wave is unchanged, however, we show that the \lq\lq ridge\rq\rq~of the
rogue wave at this time is distinctly tilted to the $t$ axis,
and its skew angle is $\arctan K^{-1}$, see Fig. \ref{fig:2}(b).
The analogous phenomena produced by the higher-order effects have been comprehensively studied for rogue waves
in the Hirota equation \cite{24},
Sasa-Satsuma equation \cite{25} and Kundu-Eckhaus equation \cite{35,38},
while for the coupled system without any higher-order terms they have been rarely reported.

Next, following the general rational solution (\ref{solu}) with $N=2$, one can end up with the
second-order rogue wave solution. Since the expression of the higher-order rogue wave
solution involving several free parameters is very cumbersome, here we only write down a special representation
of it that features fundamental pattern under the choice of $a=1,c=1,d_{0}=1,\omega=0$ and $s_{1}=0$, namely,
\begin{equation}\label{15}
E[2]=e^{i\theta}\dfrac{F_{2}+iG_{2}}{D_{2}},
\end{equation}
where $F_{2}$, $D_{2}$ and $G_{2}$ are the rational polynomials, see appendix A.

\begin{figure}[!h]
\centering
\renewcommand{\figurename}{{\bf Fig.}}
{\includegraphics[height=6cm,width=8.5cm]{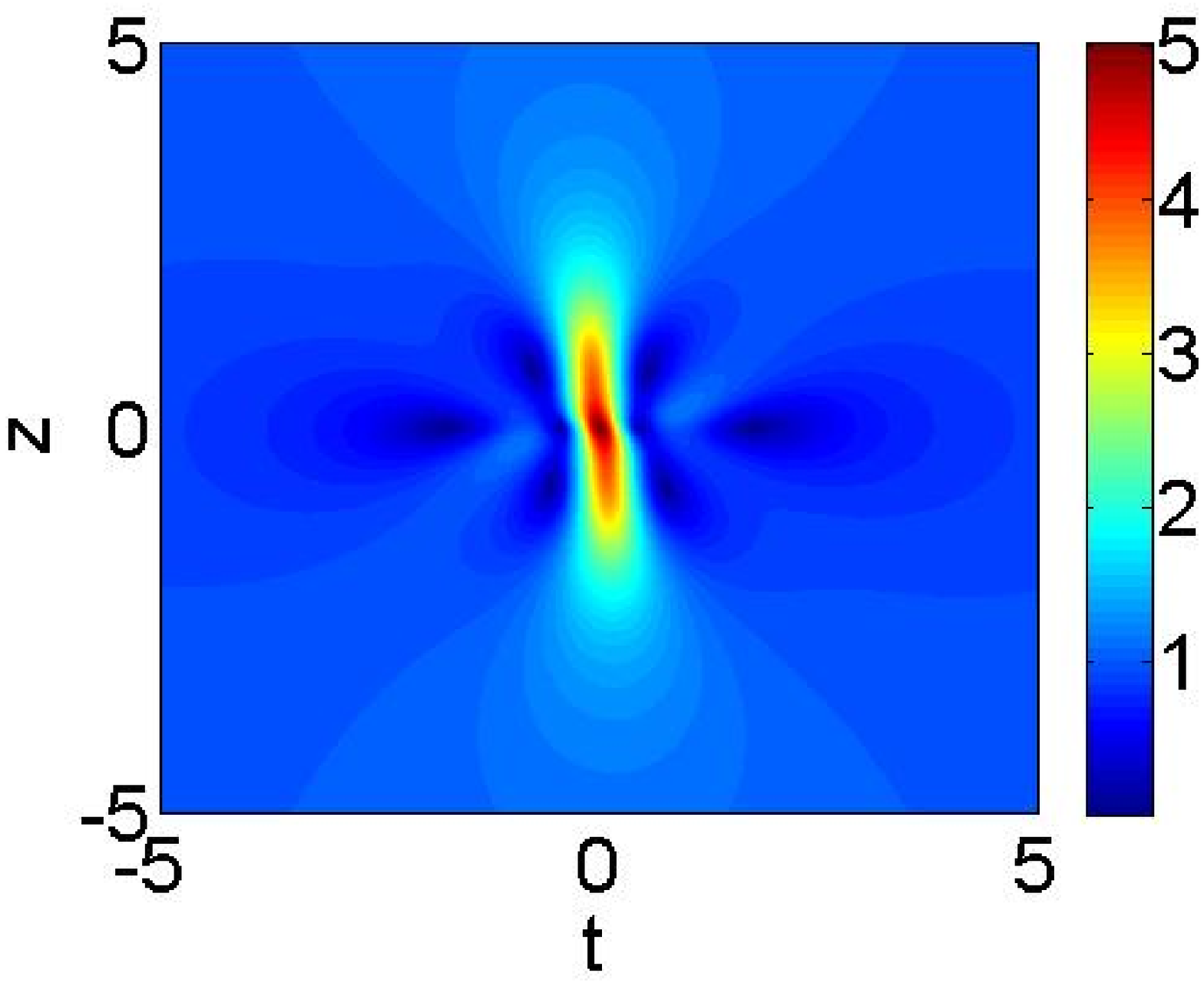}}
{\includegraphics[height=6cm,width=8.5cm]{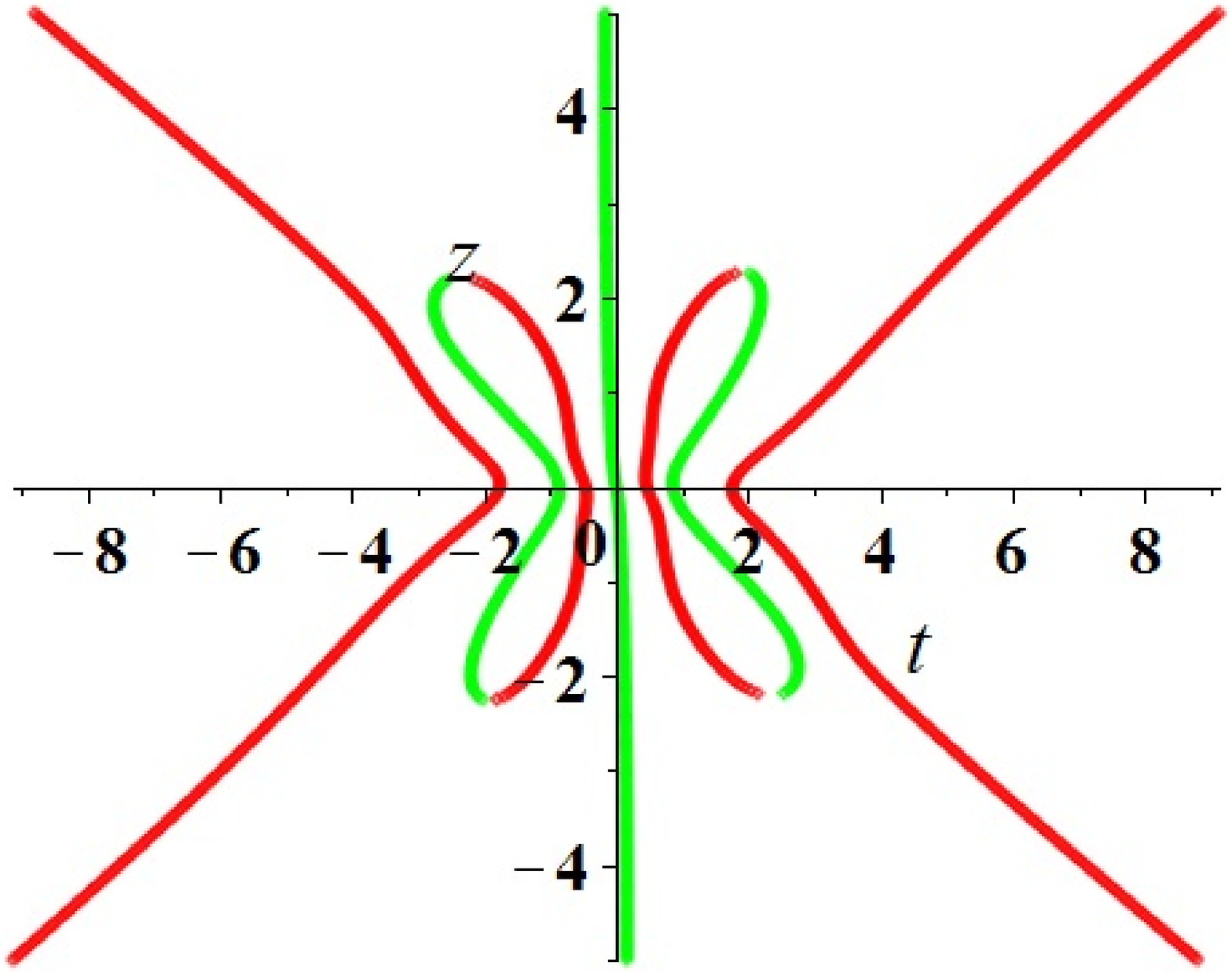}}
\begin{center}
\hskip 1cm $(\rm{a})$ \hskip 7cm $(\rm{b})$
\end{center}
\caption{(a) Density plot and (b) trace of the second-order rogue wave of fundamental pattern given by Eq. (\ref{15})
with $b=3/2$. }
\label{fig:3}
\end{figure}

\begin{figure}[!h]
\centering
\renewcommand{\figurename}{{\bf Fig.}}
{\includegraphics[height=6cm,width=8.5cm]{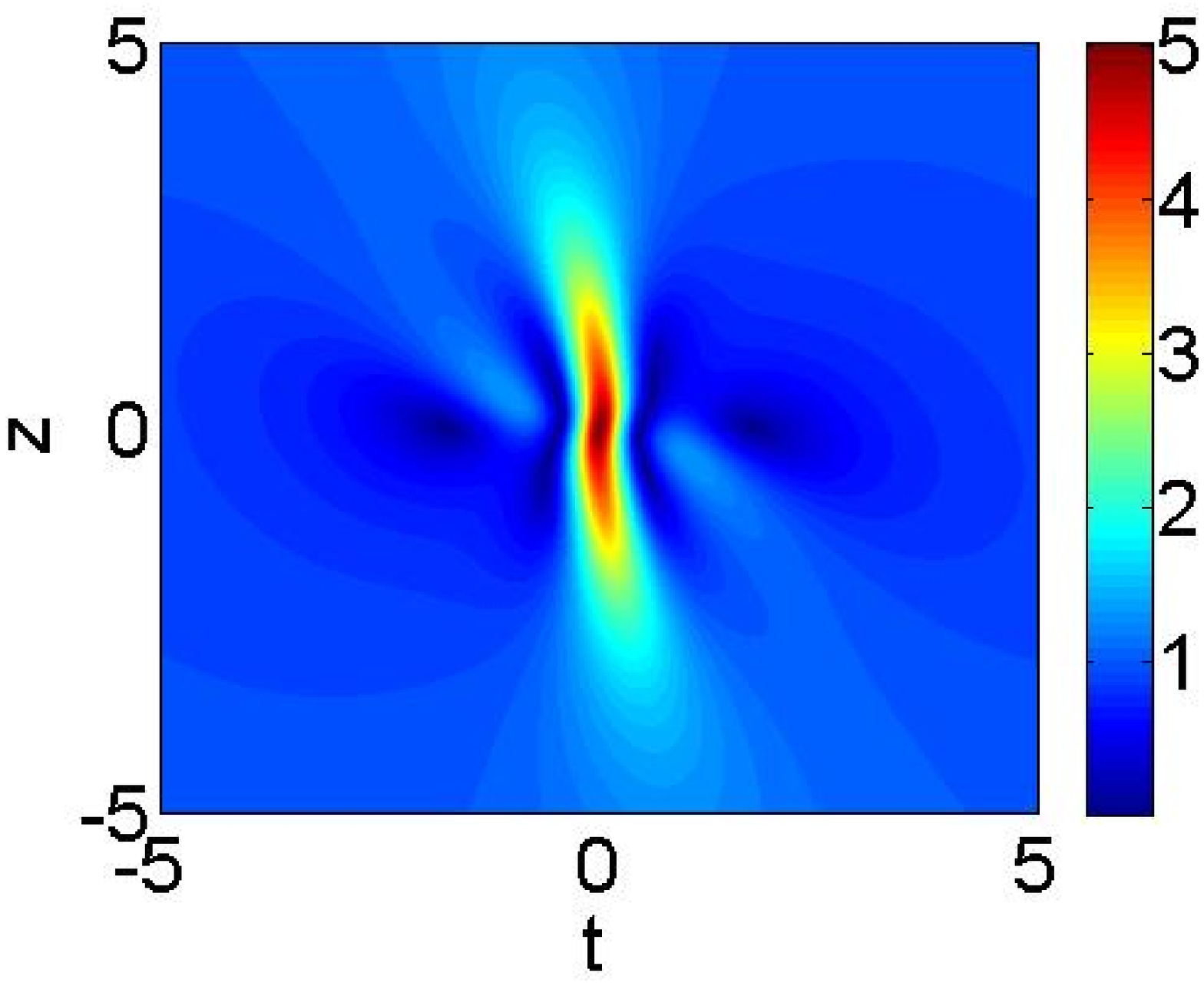}}
{\includegraphics[height=6cm,width=8.5cm]{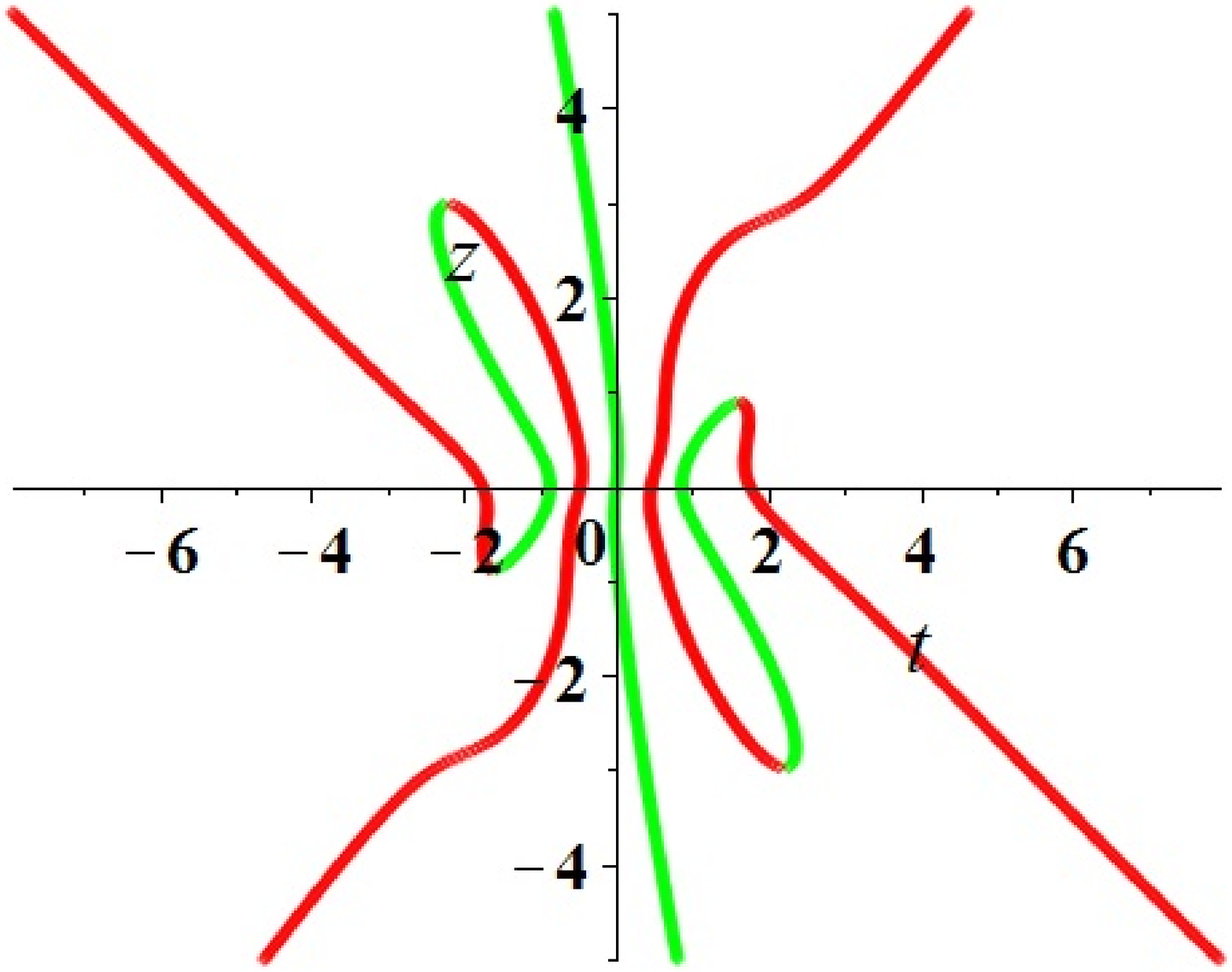}}
\begin{center}
\hskip 1cm $(\rm{a})$ \hskip 7cm $(\rm{b})$
\end{center}
\caption{(a) Density plot and (b) trace of the second-order rogue wave of fundamental pattern given by Eq. (\ref{15}) with $b=-3/2$.}
\label{fig:4}
\end{figure}

At this stage, it is apparent that there remains only one free parameter $b$ in the above solution.
The dynamics and trace characteristics of the rogue waves described by Eq. (\ref{15}) with different choices of $b$
are presented in Figs. \ref{fig:3} and \ref{fig:4}. The maximum amplitude of the second-order rogue wave is 5
and occurs at $(0,0)$.
We observe that like the first-order rogue wave in Fig. \ref{fig:2}(a),
when choosing $b=-3/2$ as an example to ensure that $K\neq0$,
the \lq\lq ridge\rq\rq~of the second-order rogue wave in Fig. \ref{fig:4}(a) is also significantly tilted to the
$t$ axis.

As a matter of fact, we notice that,  when taking $t\rightarrow\infty$ and $z\rightarrow\infty$,
the trace of the higher-order rogue wave is asymptotic to that of the first-order rogue wave under the same initial excitations.
For instance, in Figs. \ref{fig:3}(b) and \ref{fig:4}(b),
when we choose the temporal and spacial intervals large enough,
the traces shown in these two figures would just be contracted
into the \lq\lq X\rq\rq~shape given in Figs. \ref{fig:1}(b) and \ref{fig:2}(b).
Thus, the skew angle of the \lq\lq ridge\rq\rq~of the higher-order rogue wave is actually determined by
that of the first-order rogue wave under the same initial excitations, i.e. the value of $\arctan K^{-1}$.

Hereby, we need to point out that, unlike the trace of the first-order rogue wave, the expressions of the
trajectories of the humps and valleys of the second- or  higher-order rogue wave
can not be explicitly calculated, since the complicated high-order algebraic equation emerges in these extreme points.
Consequently, in our paper, we employe the numerical method to describe the traces of
the higher-order rogue waves, namely, the trajectories of
the extreme points in the expressions of the higher-order rogue wave solutions.

\begin{figure}[!h]
\centering
\renewcommand{\figurename}{{\bf Fig.}}
{\includegraphics[height=6cm,width=8.5cm]{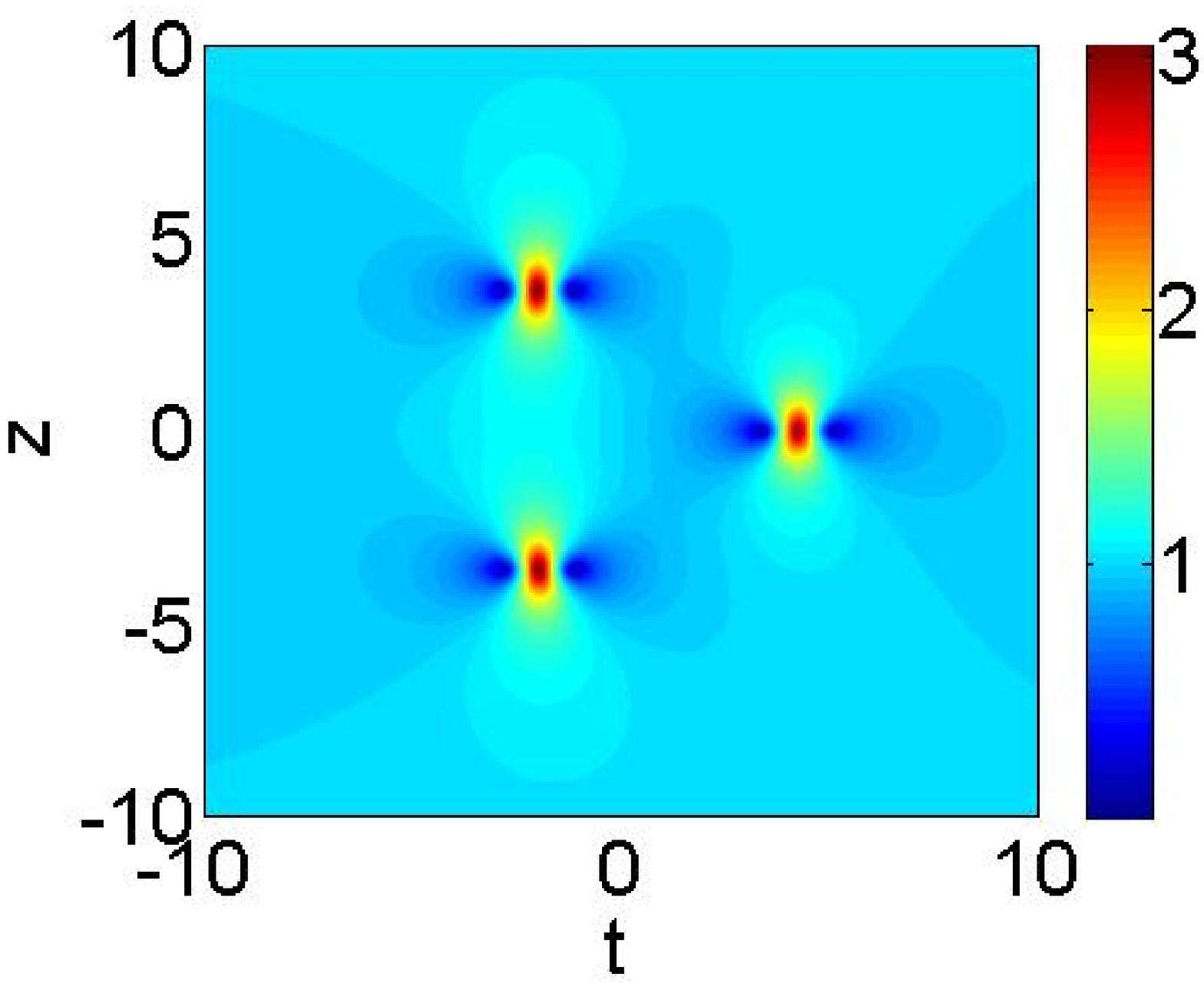}}
{\includegraphics[height=6cm,width=8.5cm]{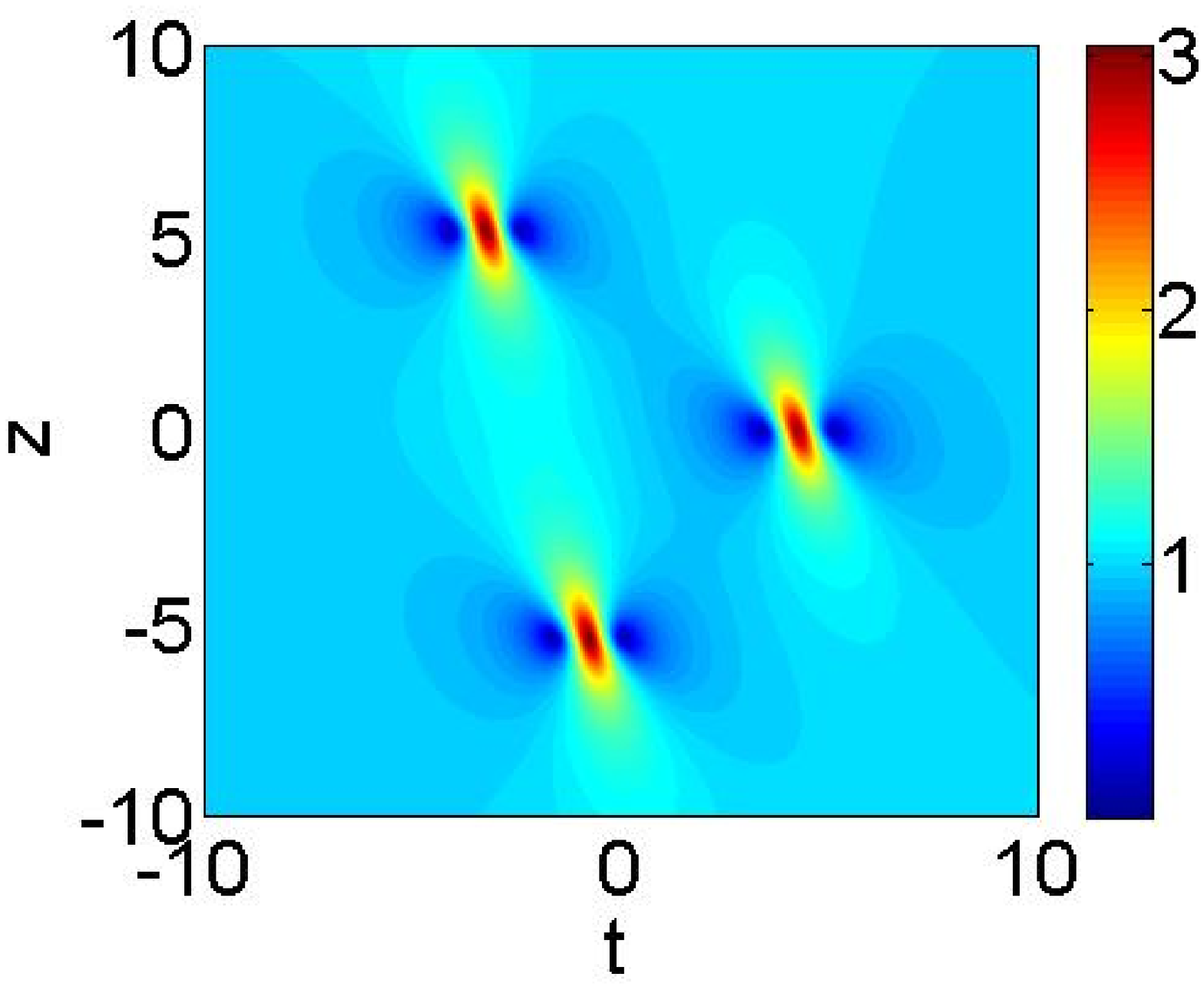}}
\begin{center}
\hskip 1cm $(\rm{a})$ \hskip 7cm $(\rm{b})$
\end{center}
\caption{(a), (b) Density plots of the second-order rogue waves of triangular pattern by choosing
$b=3/2$ and $b=-3/2$ with $a=1,c=1,d_{0}=1,\omega=0,s_{1}=100$. }
\label{fig:5}
\end{figure}

On the other hand, when letting $s_{1}\neq0$, the fundamental second-order rogue wave can be separated into three
first-order rogue waves distributing with a triangle, see Fig. \ref{fig:5}.
As is confirmed in Fig. \ref{fig:5}(b), when choosing the value of $b$ satisfying $K\neq0$,
the \lq\lq ridge\rq\rq~of each of the rogue waves is tilted to the $t$ axis.
Additionally, one can also find that rogue waves in Fig. \ref{fig:5}(b) are expanded in the
$z$ dimension, the humps in Fig. \ref{fig:5}(a) are localized at $(-2.0173,3.6432)$, $(4.2172,0)$ and
$(-1.9832,-3.5864)$,  while in Fig. \ref{fig:5}(b), the corresponding coordinates
become $(-3.2514,5.1937)$, $(4.2172,0)$ and $(-0.7858,-5.3658)$.

Afterwards, by choosing the adequate free parameters $s_{1}$ and $s_{2}$,
we can show the evolution plots of the third-order rogue waves of fundamental,
triangular and circular patterns, respectively.
Here, we completely refrain from giving the concrete expressions of the burdensome rational polynomials,
which can be directly constructed by means of Eq. (\ref{solu}) with $N=3$.

\begin{figure}[!h]
\centering
\renewcommand{\figurename}{{\bf Fig.}}
{\includegraphics[height=6cm,width=8.5cm]{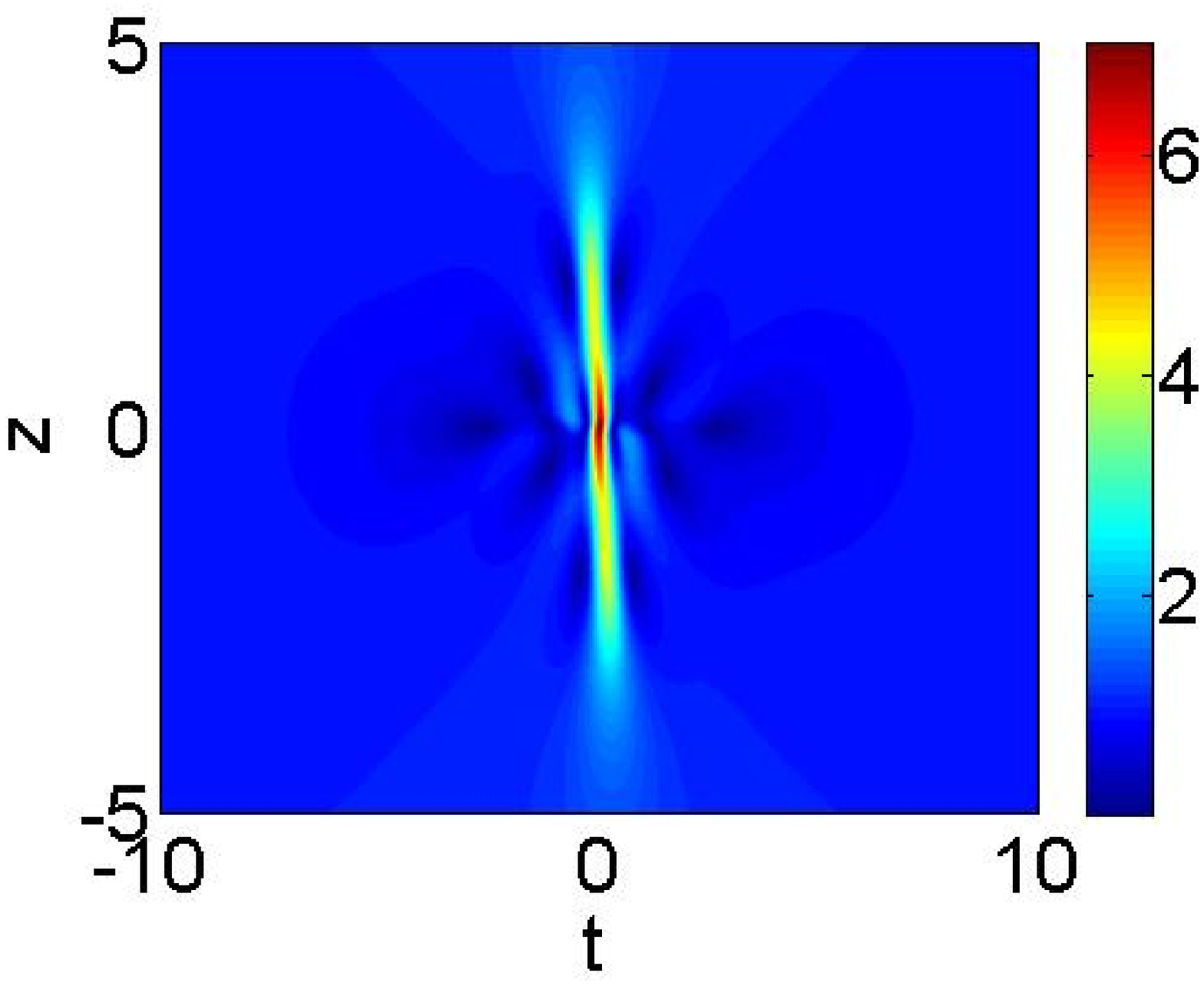}}
{\includegraphics[height=6cm,width=8.5cm]{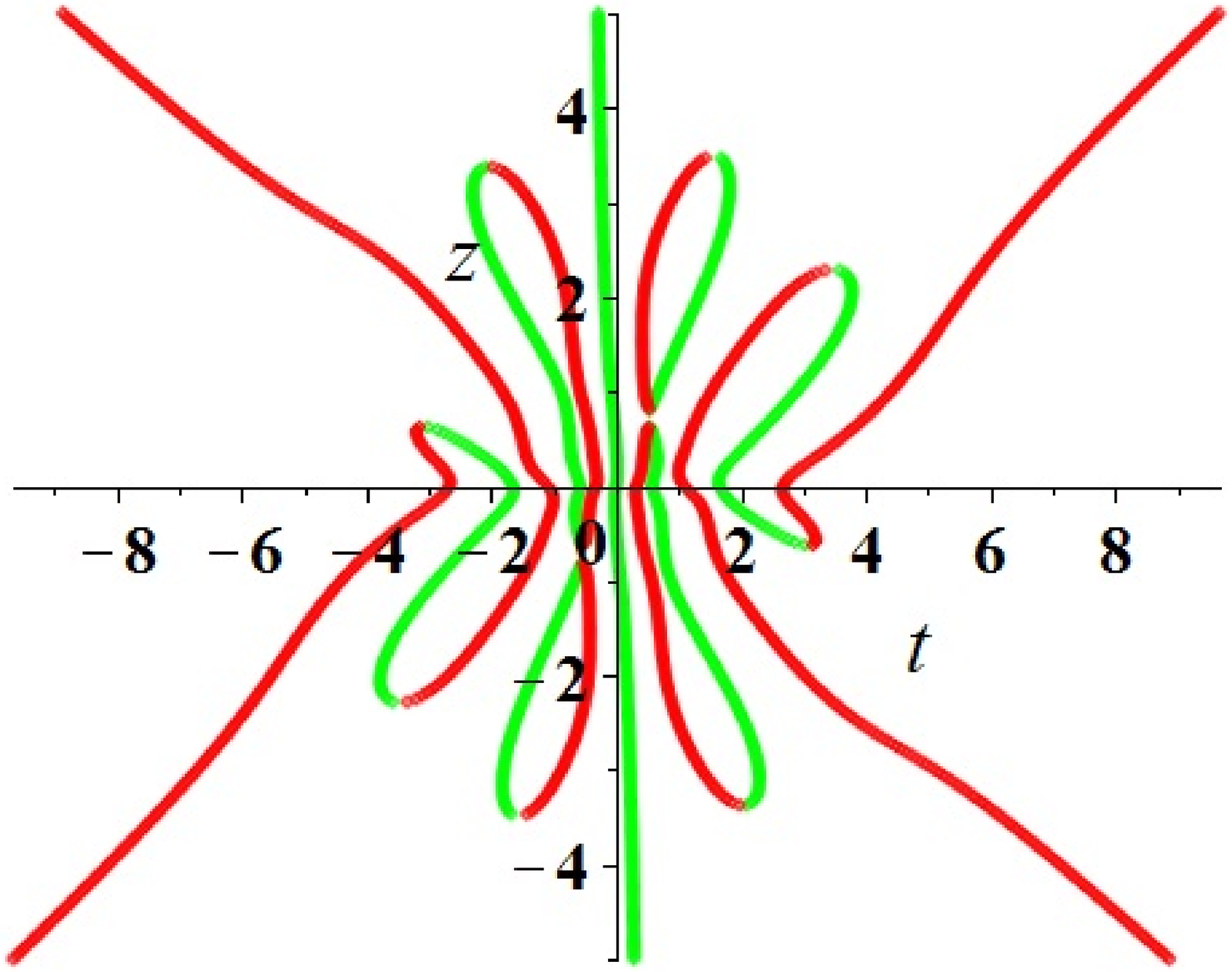}}
\begin{center}
\hskip 1cm $(\rm{a})$ \hskip 7cm $(\rm{b})$
\end{center}
\caption{(a) Density plot and (b) trace of the third-order rogue wave of fundamental pattern
with $a=1,b=3/2,c=1,d_{0}=1,\omega=0,s_{1}=0,s_{2}=0$. }
\label{fig:6}
\end{figure}

\begin{figure}[!h]
\centering
\renewcommand{\figurename}{{\bf Fig.}}
{\includegraphics[height=6cm,width=8.5cm]{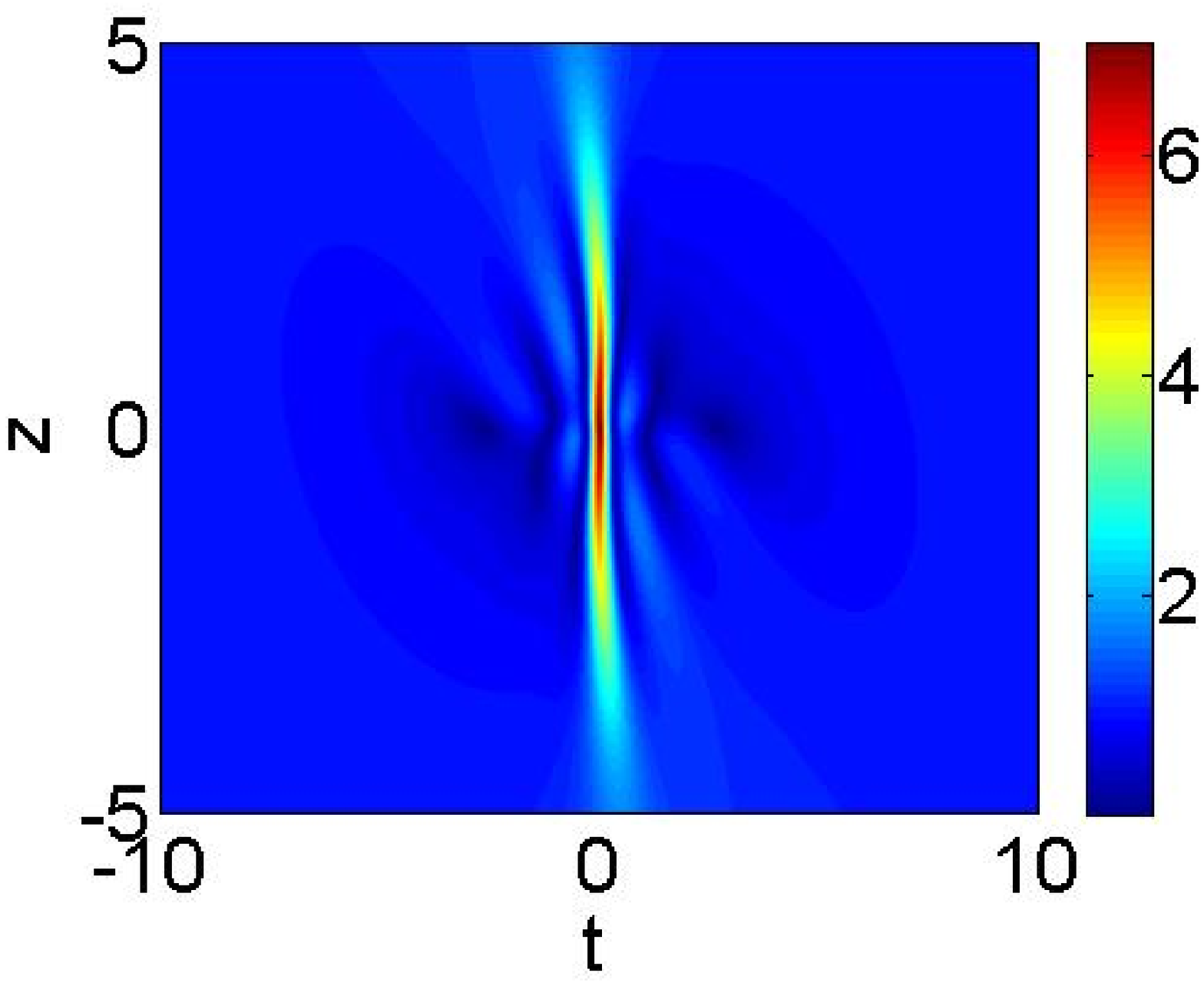}}
{\includegraphics[height=6cm,width=8.5cm]{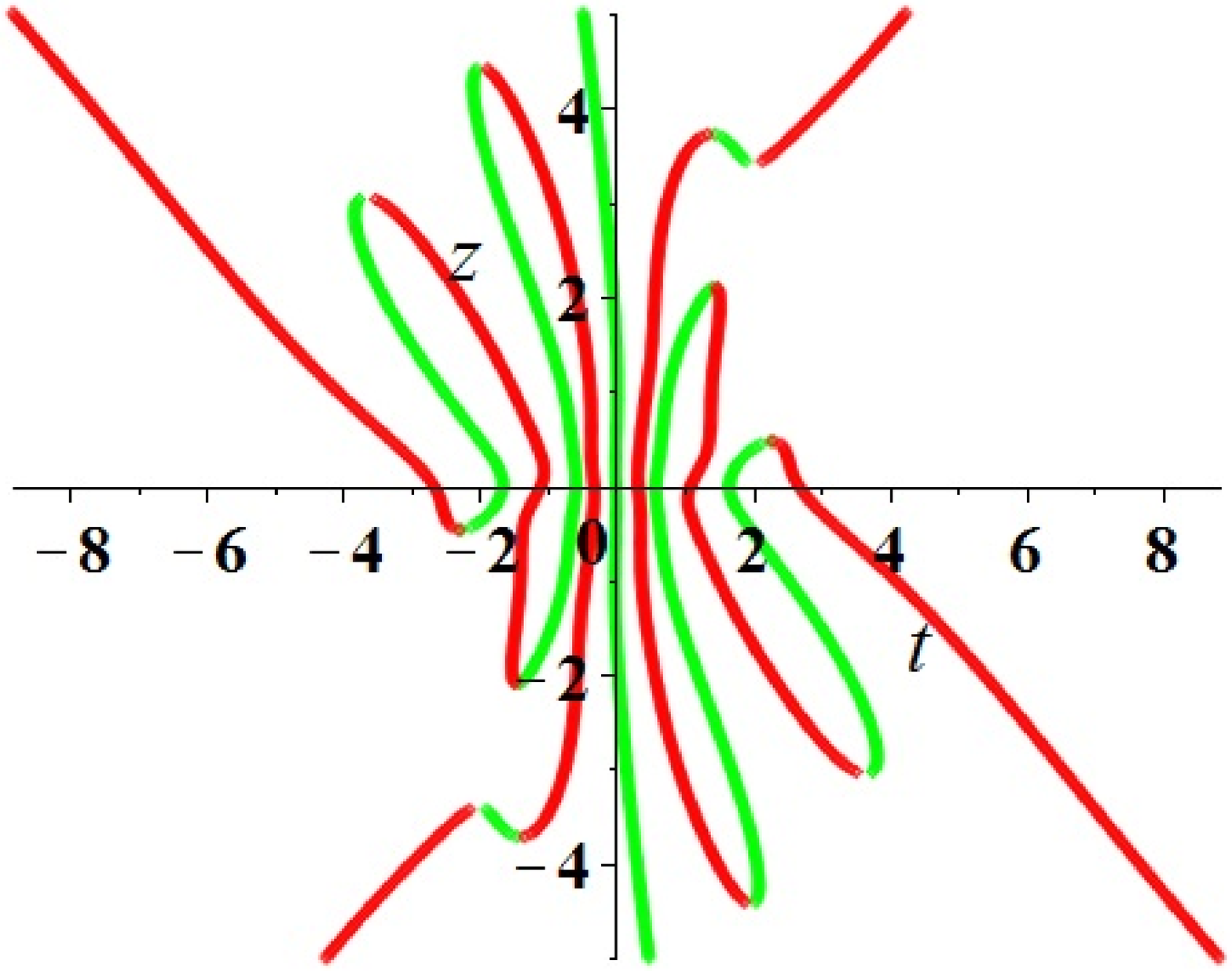}}
\begin{center}
\hskip 1cm $(\rm{a})$ \hskip 7cm $(\rm{b})$
\end{center}
\caption{(a) Density plot and (b) trace of the third-order rogue wave of fundamental pattern
with $a=1,b=-3/2,c=1,d_{0}=1,\omega=0,s_{1}=0,s_{2}=0$. }
\label{fig:7}
\end{figure}

\begin{figure}[!h]
\centering
\renewcommand{\figurename}{{\bf Fig.}}
{\includegraphics[height=6cm,width=8.5cm]{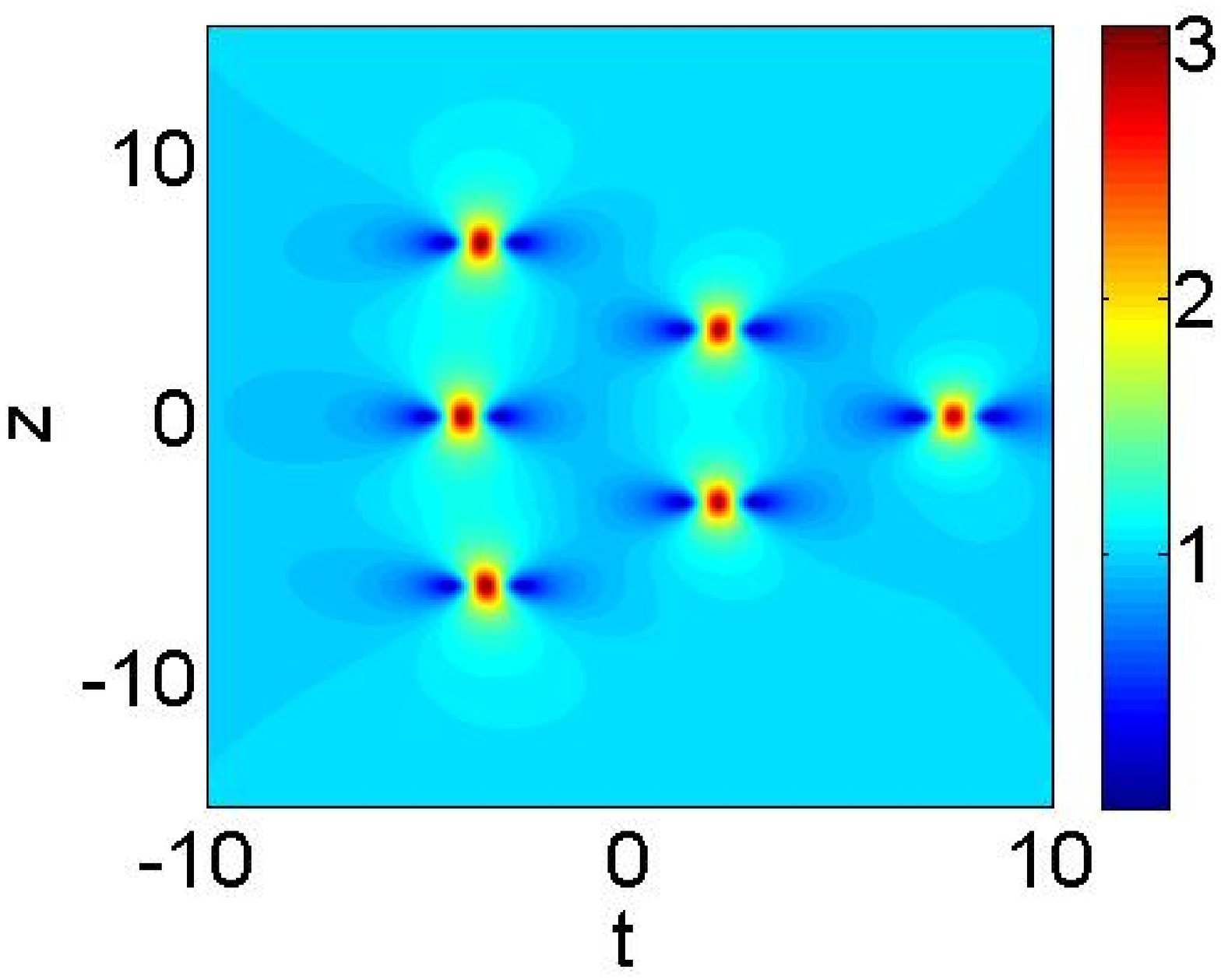}}
{\includegraphics[height=6cm,width=8.5cm]{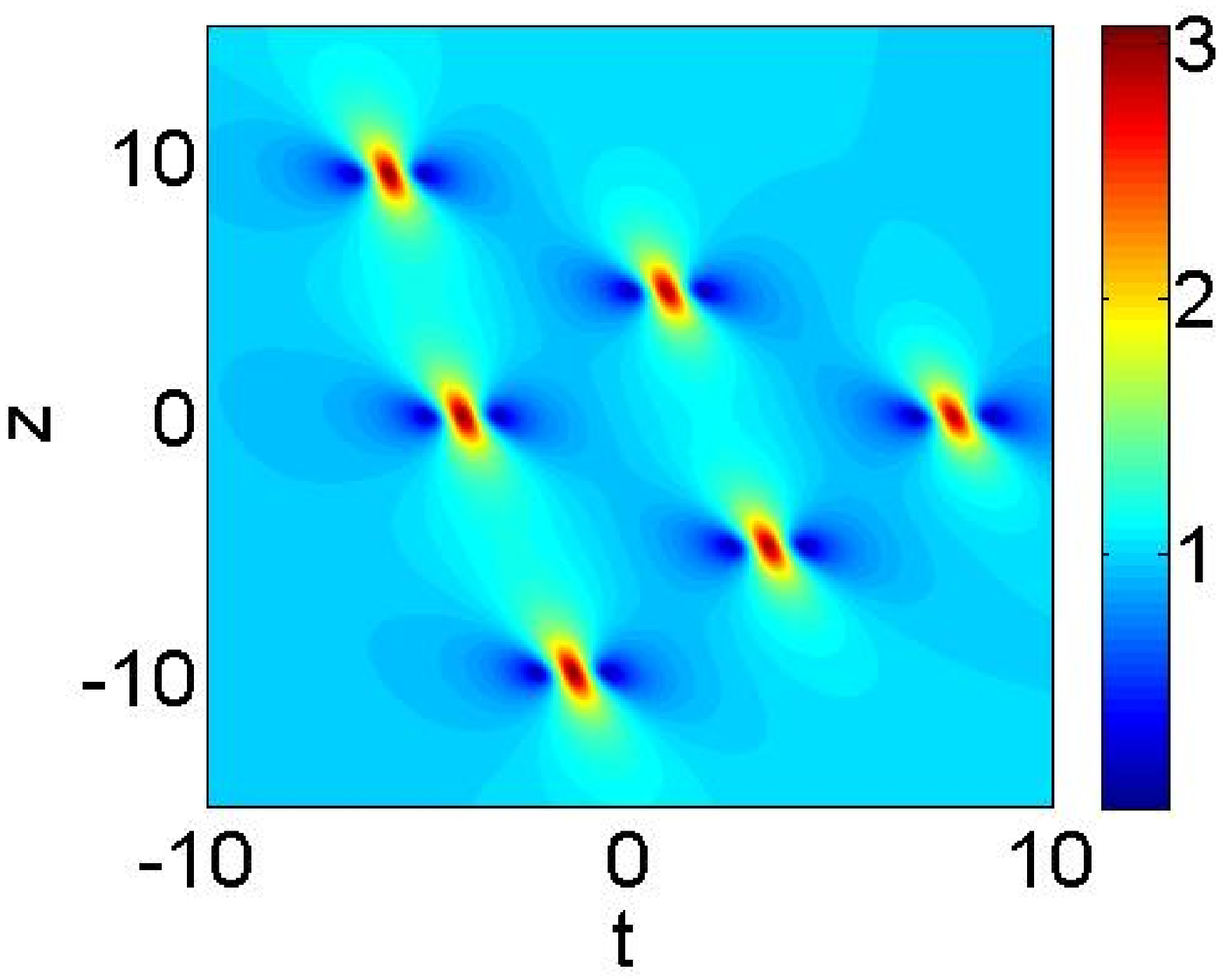}}
\begin{center}
\hskip 1cm $(\rm{a})$ \hskip 7cm $(\rm{b})$
\end{center}
\caption{(a), (b) Density plots of the third-order rogue waves of triangular pattern by choosing
$b=3/2$ and $b=-3/2$ with $a=1,c=1,d_{0}=1,\omega=0,s_{1}=100,s_{2}=0$. }
\label{fig:8}
\end{figure}

\begin{figure}[!h]
\centering
\renewcommand{\figurename}{{\bf Fig.}}
{\includegraphics[height=6cm,width=8.5cm]{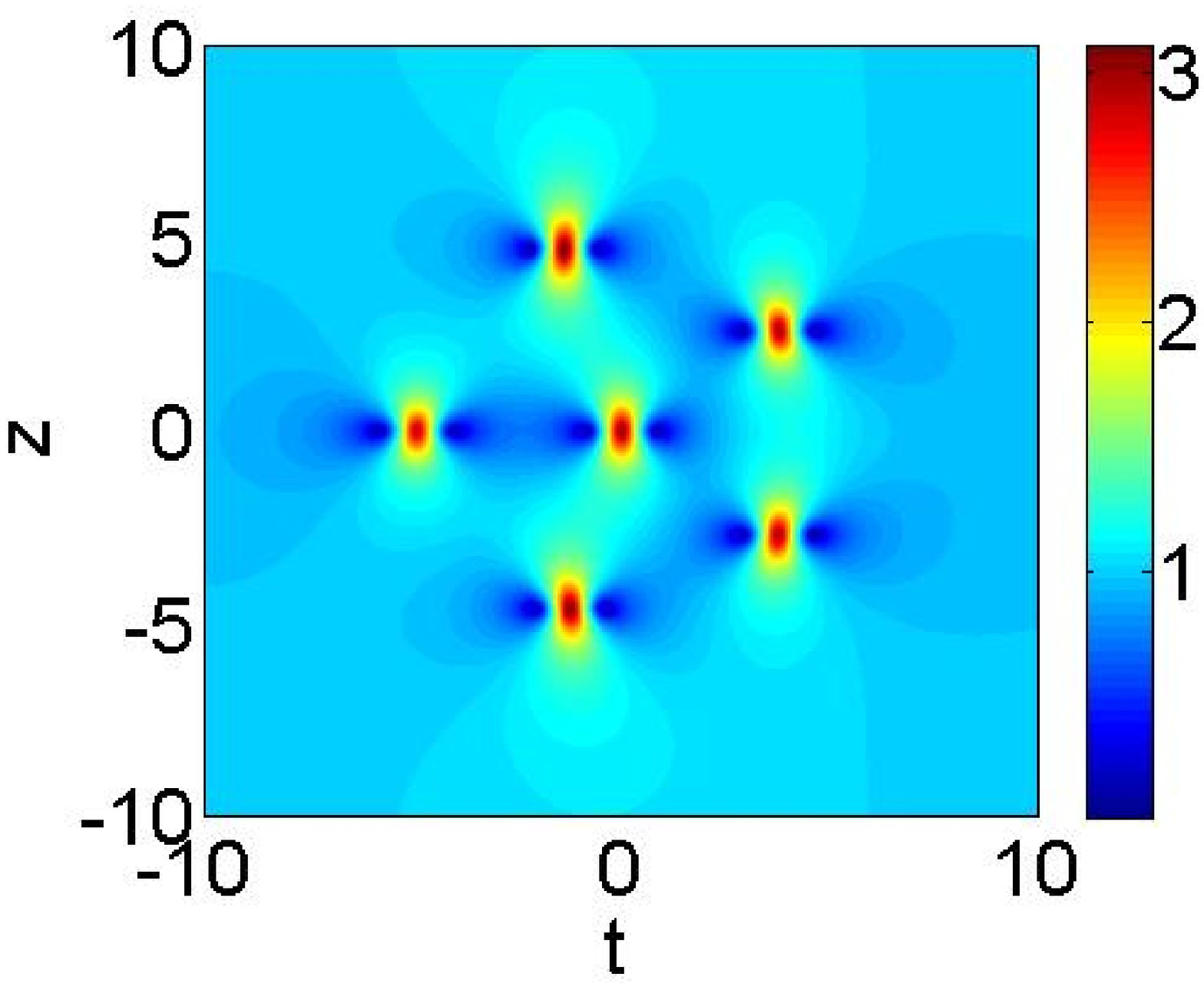}}
{\includegraphics[height=6cm,width=8.5cm]{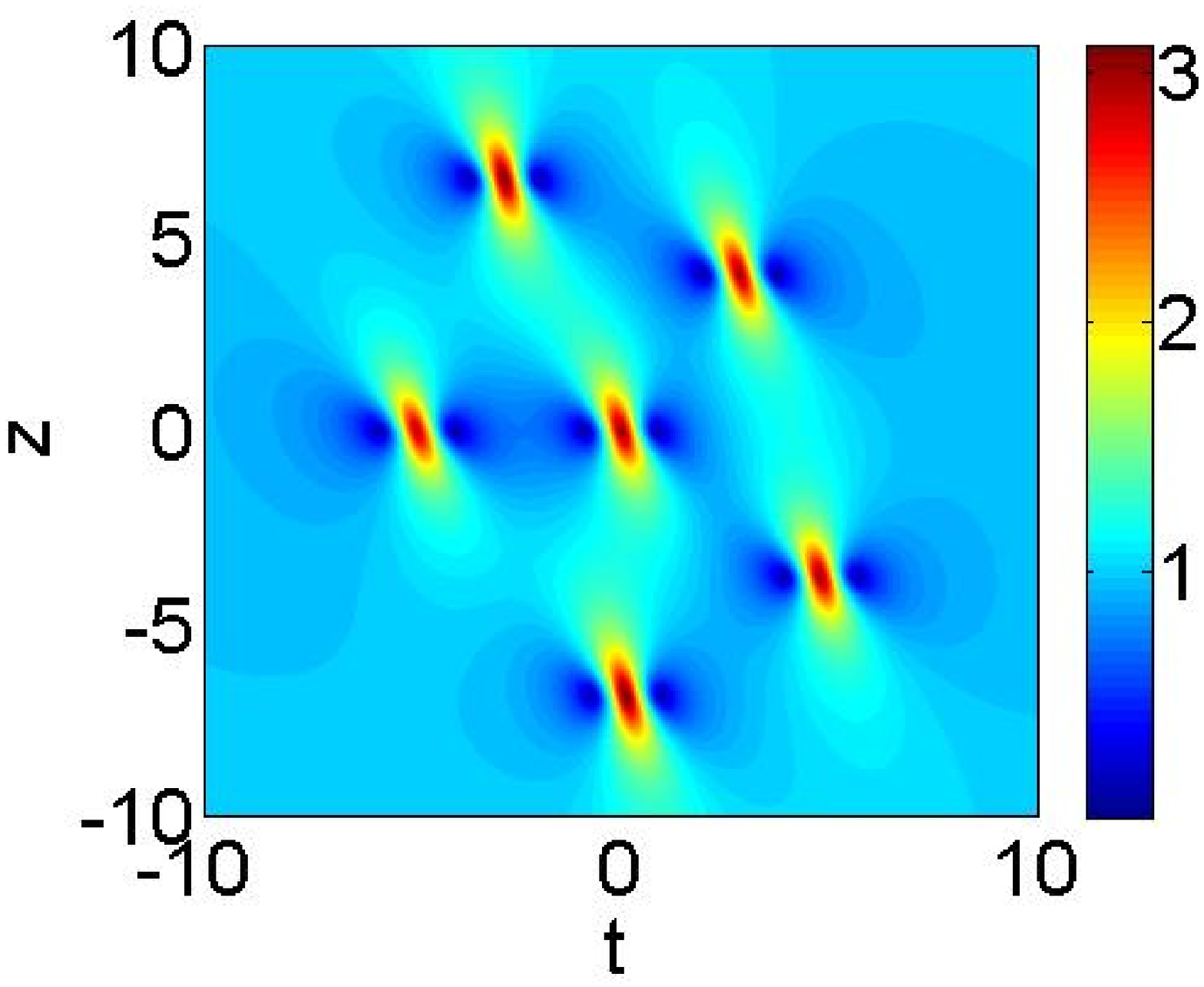}}
\begin{center}
\hskip 1cm $(\rm{a})$ \hskip 7cm $(\rm{b})$
\end{center}
\caption{(a), (b) Density plots of the third-order rogue waves of circular pattern by choosing
$b=3/2$ and $b=-3/2$ with $a=1,c=1,d_{0}=1,\omega=0,s_{1}=0,s_{2}=1000$. }
\label{fig:9}
\end{figure}

As are depicted in Figs. \ref{fig:6} and \ref{fig:7}, by taking $s_{1}=0$ and $s_{2}=0$, the third-order rogue waves of
fundamental pattern are shown. The maximum amplitude of the third-order rogue wave is 7
and appears at $(0,0)$.
Meanwhile, when selecting $s_{1}\neq0,s_{2}=0$ or $s_{1}=0,s_{2}\neq0$,
the fundamental third-order rogue wave can split into six first-order rogue waves arraying with a triangle or a ring,
respectively, see Figs. \ref{fig:8} and \ref{fig:9}.
Likewise, it is clearly observed that the \lq\lq ridge\rq\rq~of each of
the rogue waves in Figs. \ref{fig:7}(a), \ref{fig:8}(b) and \ref{fig:9}(b)
with $b=-3/2$ is tilted to the $t$ axis in contrast with the same initial excitations except $b=3/2$, and the range of the
appearance for these rogue waves in the $z$ dimension is notablely expanded.
For example, the maximum amplitudes of the six first-order rogue waves in Fig. \ref{fig:8}(a) are
localized at $(-3.5291,6.6662)$, $(2.0912,3.3385)$, $(-3.9545,0)$, $(7.6248,0)$,
$(2.0701,-3.2792)$ and $(-3.4100,-6.4934)$, while in Fig. \ref{fig:8}(b), the rogue waves are distinctly expanded in the
$z$ dimension, the corresponding coordinates turn into
$(-5.7003,9.2761)$, $(0.8638,4.8039)$, $(-3.9545,0)$, $(7.6248,0)$,
$(3.2612,-4.9716)$ and $(-1.3502,-9.8189)$.

\subsection{W-shaped soliton}

As is announced before, when letting the MI growth rate tend to zero under the zero-frequency MS region, one can
bring about the state transition from rogue wave to W-shaped soliton. For that purpose, we take
\begin{equation}\label{16}
b= b_{s}|_{\kappa=0}=\dfrac{2(2\omega-a)d_{0}}{3 a^2-12a\omega+4c^2+12\omega^2 }.
\end{equation}
Then through the direct substitution of Eq. (\ref{16}) into Eq. (\ref{R1}),
we can define the motions of the first-order W-shaped soliton\rq~hump and valleys analytically
\begin{equation}\label{ths}
T_{h}^{'}=-K^{'}z,
\end{equation}
and
\begin{equation}\label{tvs}
T_{v}^{'}=-K^{'}z\pm\frac{\sqrt{3}}{2},
\end{equation}
where
$$K^{'}=\dfrac{d_{0}}{3a^2-12a\omega+4c^2+12\omega^2}.$$

\begin{figure}[!h]
\centering
\renewcommand{\figurename}{{\bf Fig.}}
{\includegraphics[height=6cm,width=8.5cm]{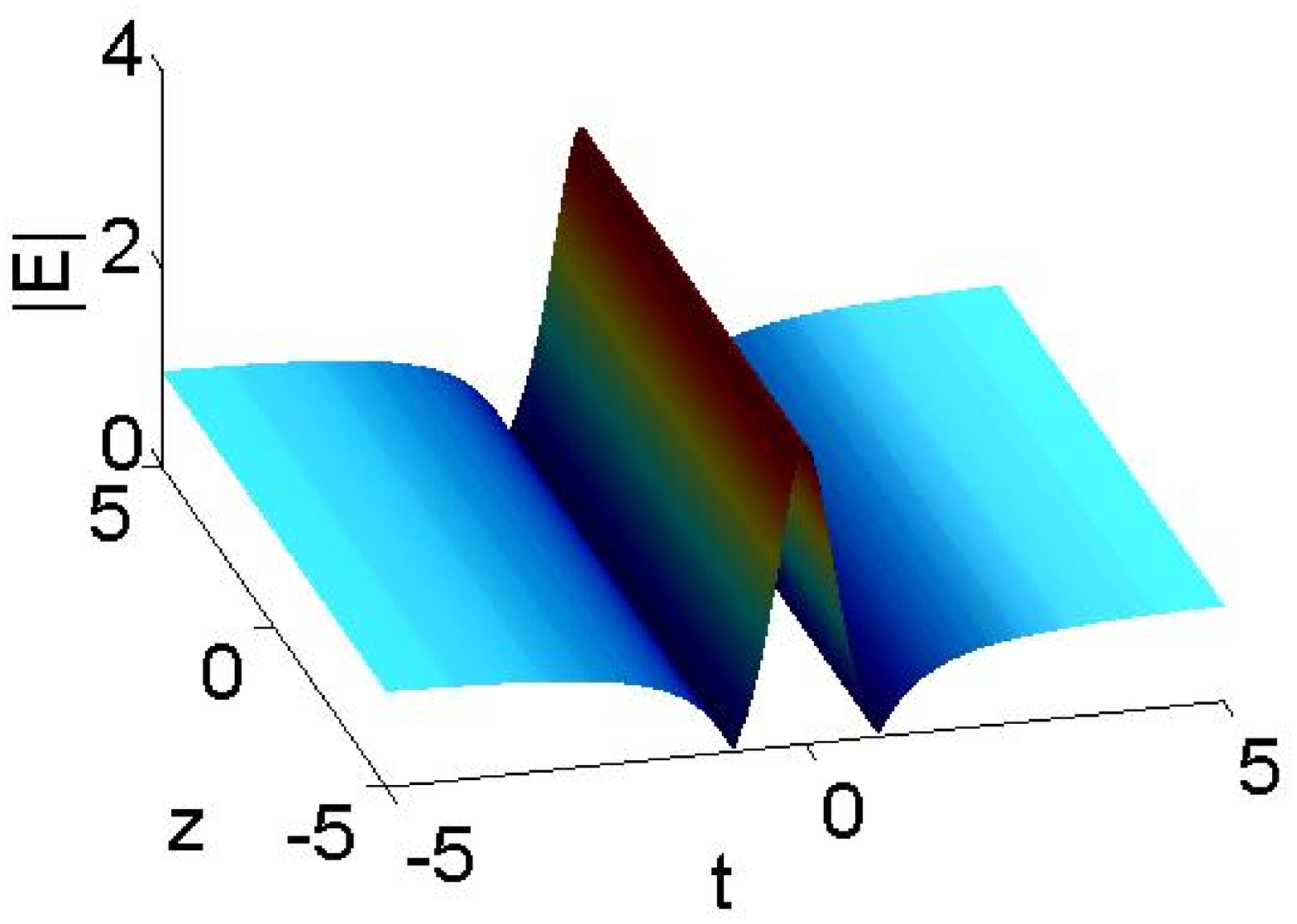}}
{\includegraphics[height=6cm,width=8.5cm]{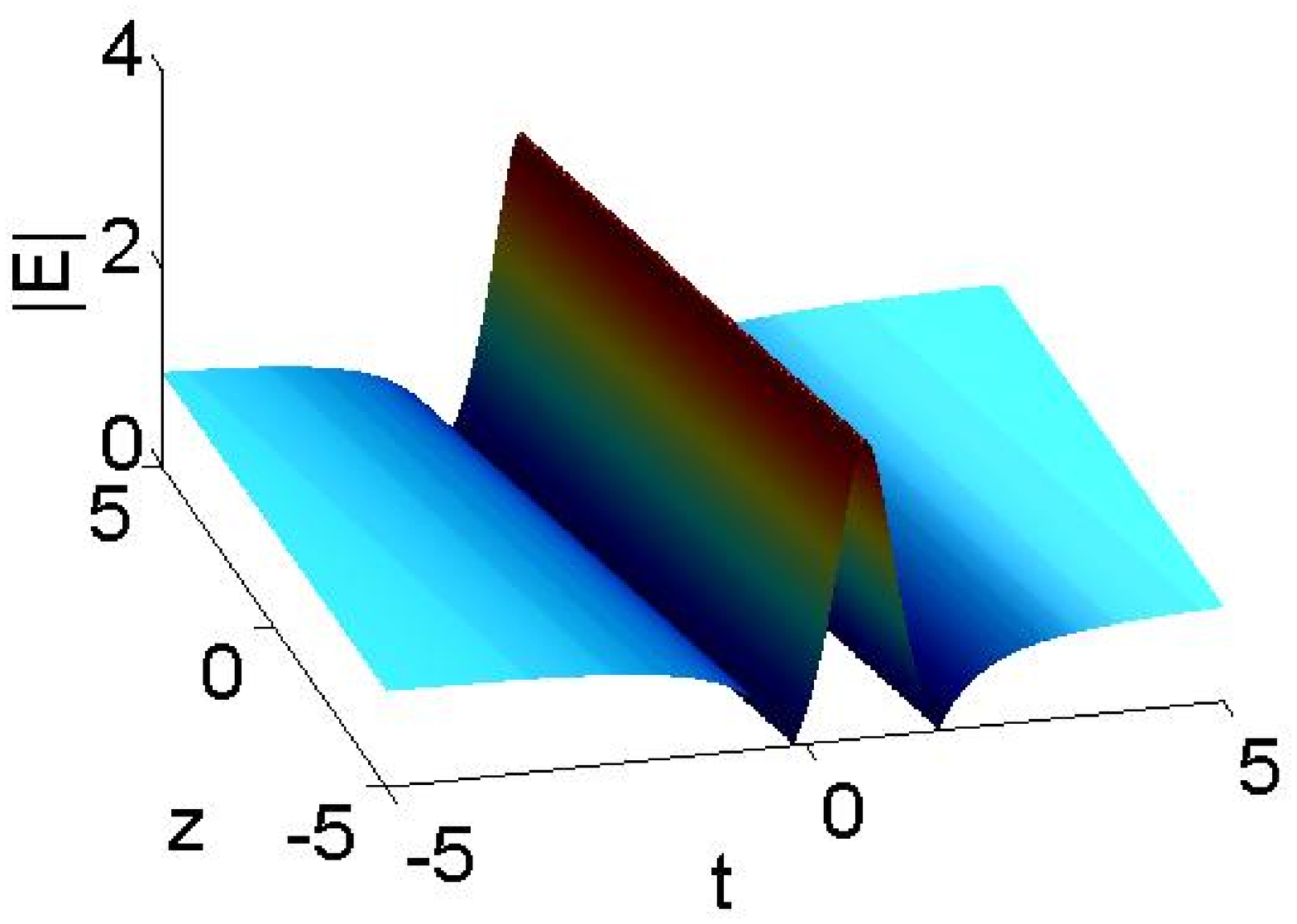}}
\begin{center}
\hskip 1cm $(\rm{a})$ \hskip 7cm $(\rm{b})$
\end{center}
\caption{(a), (b) Density plots of the first-order stationary and nonstationary W-shaped solitons given by Eq. (\ref{R1})
with $d_{0}=0,b=0,a=1,c=1,\omega=0$ and $d_{0}=1,b=-2/7,a=1,c=1,\omega=0$.}
\label{fig:10}
\end{figure}

Here, it is facile to verify that the maximum amplitude of the first-order W-shaped soliton is 3 and
arrives at $t+K^{'}z=0$, and the minimum value of it is 0 and occurs at $t+K^{'}z=\pm\sqrt{3}/2$.
By solving $K^{'}=0$ we get $d_{0}=0$, then in this circumstance,
the trajectories of the hump and that of the valleys of the W-shaped soliton are stationary in the $z$ dimension.
In other words, the explicit soliton solution and its trace are independent on the temporal variable $z$.
At this point, we call this type of the soliton as the stationary W-shaped soliton, see Fig. \ref{fig:10}(a).
By contrast, if $d_{0}\neq0$, the explicit soliton solution dependents on both $t$ and $z$,
and its trace is determined by Eqs. (\ref{ths}) and (\ref{tvs}).
The soliton in this circumstance is viewed as the nonstationary W-shaped soliton, see Fig. \ref{fig:10}(b).

Particularly, we need to remark that, the nonstationary W-shaped soliton can only exist in the multiple SIT system (\ref{02}).
For the single SIT system (\ref{01}),
the reduction condition $e\rightarrow0$, $M\rightarrow0$ allows
$$b=\frac{d_{0}}{2\omega-a}$$
in the general plane-wave solution (\ref{07}).
By taking into account of the above equation alongside with Eq. (\ref{16}) we have $d_{0}=0$.
Thus, there only exists the stationary W-shaped soliton in system (\ref{01}),
whereas for the nonstationary one it is impossible to emerge.
The stationary W-shaped solitons in system (\ref{01}) have been recently studied
by Wang \cite{14}, and the above fact can also be readily proved through the direct MI analysis of
system (\ref{01}).

Next, under the special choice of $a=1,c=1,\omega=0$ and $s_{1}=0$, we can employe Eq. (\ref{solu}) with $N=2$ and Eq. (\ref{16}) to
obtain a second-order W-shaped soliton solution containing the important free parameter $d_{0}$, viz.
\begin{equation}\label{20}
E[2]^{'}=e^{i\theta}\dfrac{F_{2}^{'}+iG_{2}^{'}}{D_{2}^{'}},
\end{equation}
where the corresponding polynomials are explicitly given in appendix B.

\begin{figure}[!h]
\centering
\renewcommand{\figurename}{{\bf Fig.}}
{\includegraphics[height=6cm,width=8.5cm]{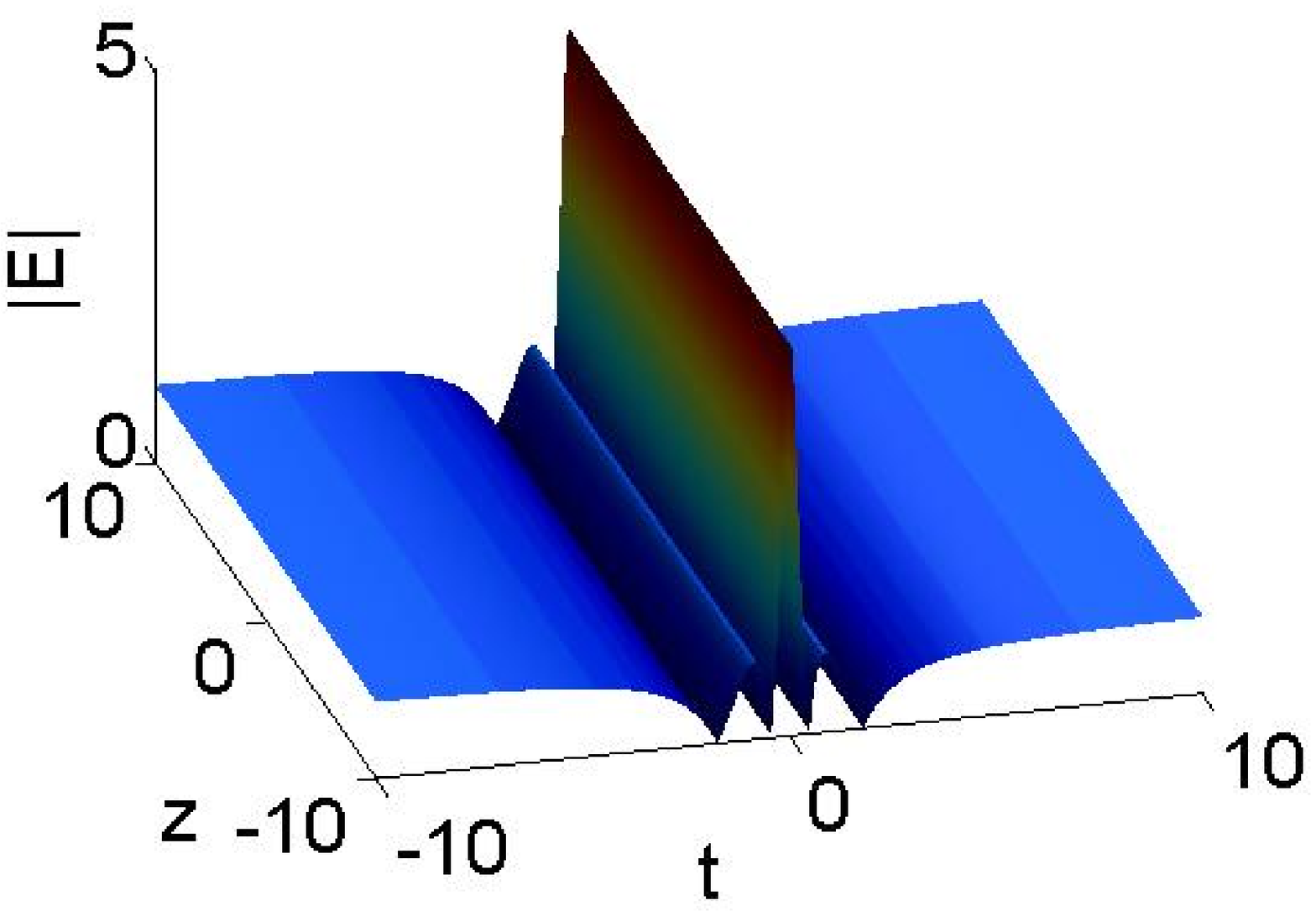}}
{\includegraphics[height=6cm,width=8.5cm]{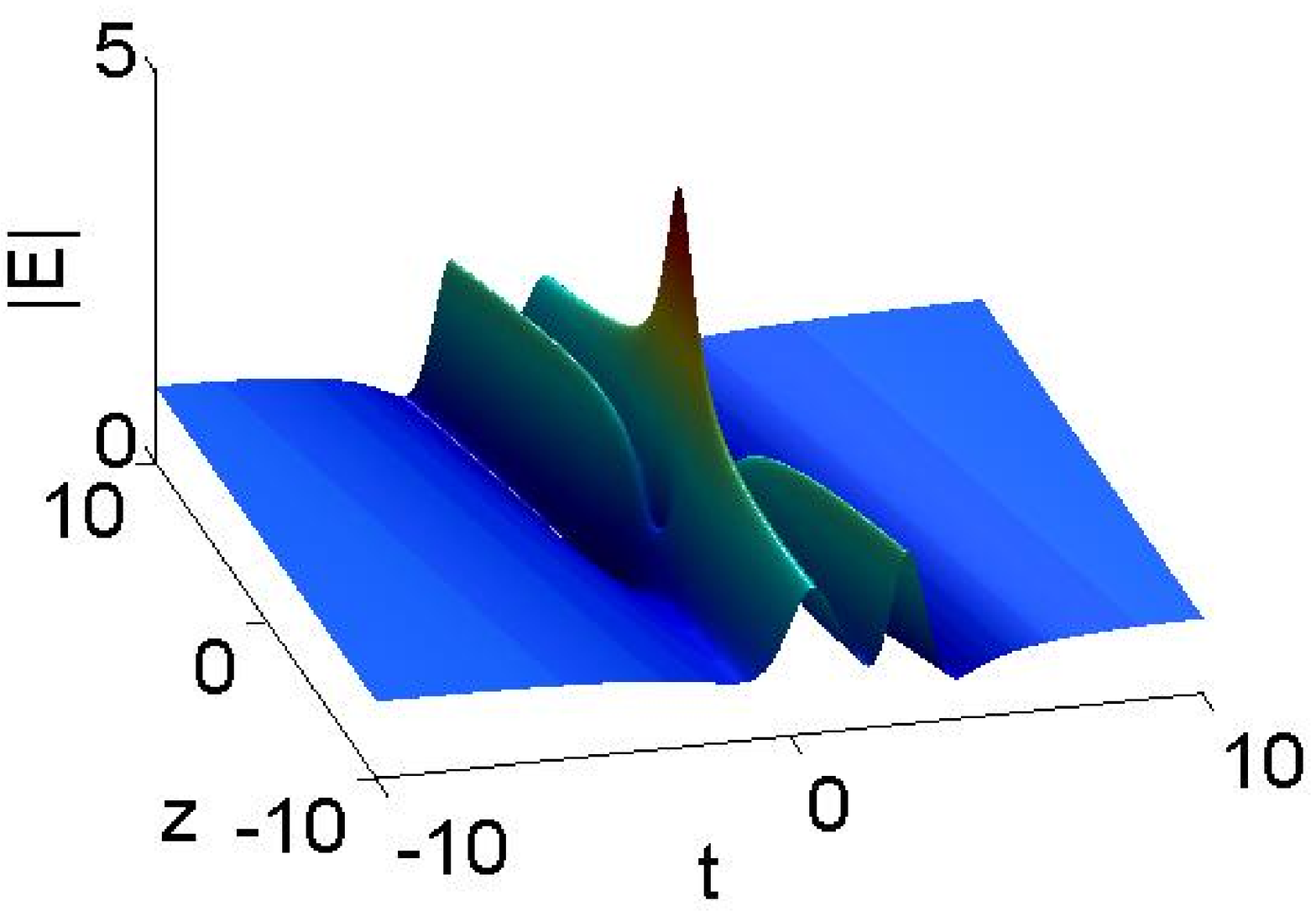}}
\begin{center}
\hskip 1cm $(\rm{a})$ \hskip 7cm $(\rm{b})$
\end{center}
\caption{(a), (b) Density plots of the second-order stationary and nonstationary W-shaped solitons gievn
by Eq. (\ref{20}) with $d_{0}=0$ and $d_{0}=1$.}
\label{fig:11}
\end{figure}

By taking $d_{0}=0$, as is exhibited in Fig. \ref{fig:11}(a),  there are three humps in the temporal-spacial plane,
the corresponding amplitudes are 5 and 1, and occur at $t=0$ and $t=\pm0.8660$.
The number of the zero-amplitude valleys is four and they arrive at $t=\pm1.7571$ and $t=\pm0.4650$.
For the nonstationary case of $d_{0}\neq0$, we display that in Fig. \ref{fig:11}(b),
two first-order nonstationary W-shaped solitons interacted with each other, and
the maximum amplitude of the second-order W-shaped soliton is 5 and is localized at $(0,0)$.

\begin{figure}[!h]
\centering
\renewcommand{\figurename}{{\bf Fig.}}
{\includegraphics[height=6cm,width=8.5cm]{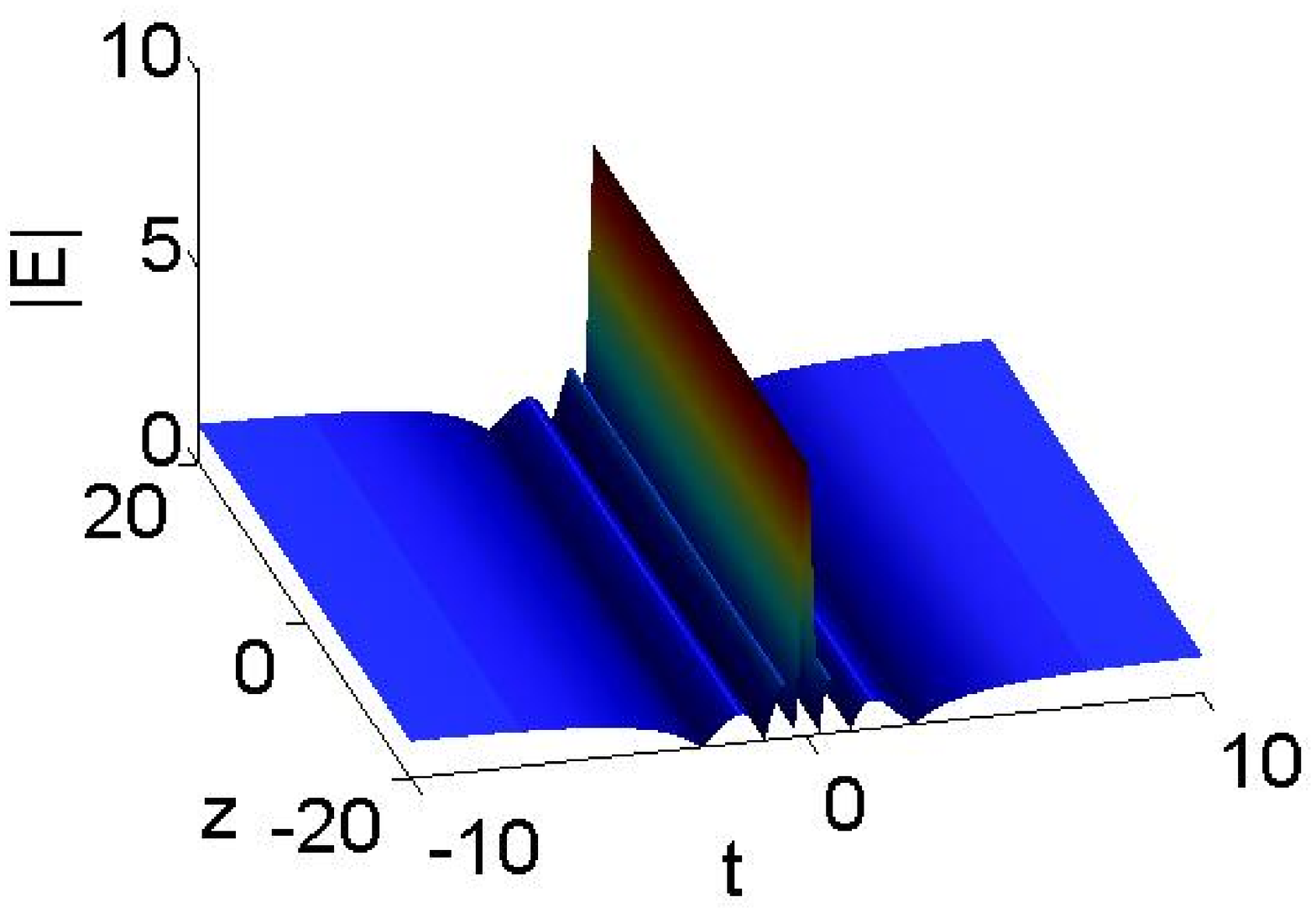}}
{\includegraphics[height=6cm,width=8.5cm]{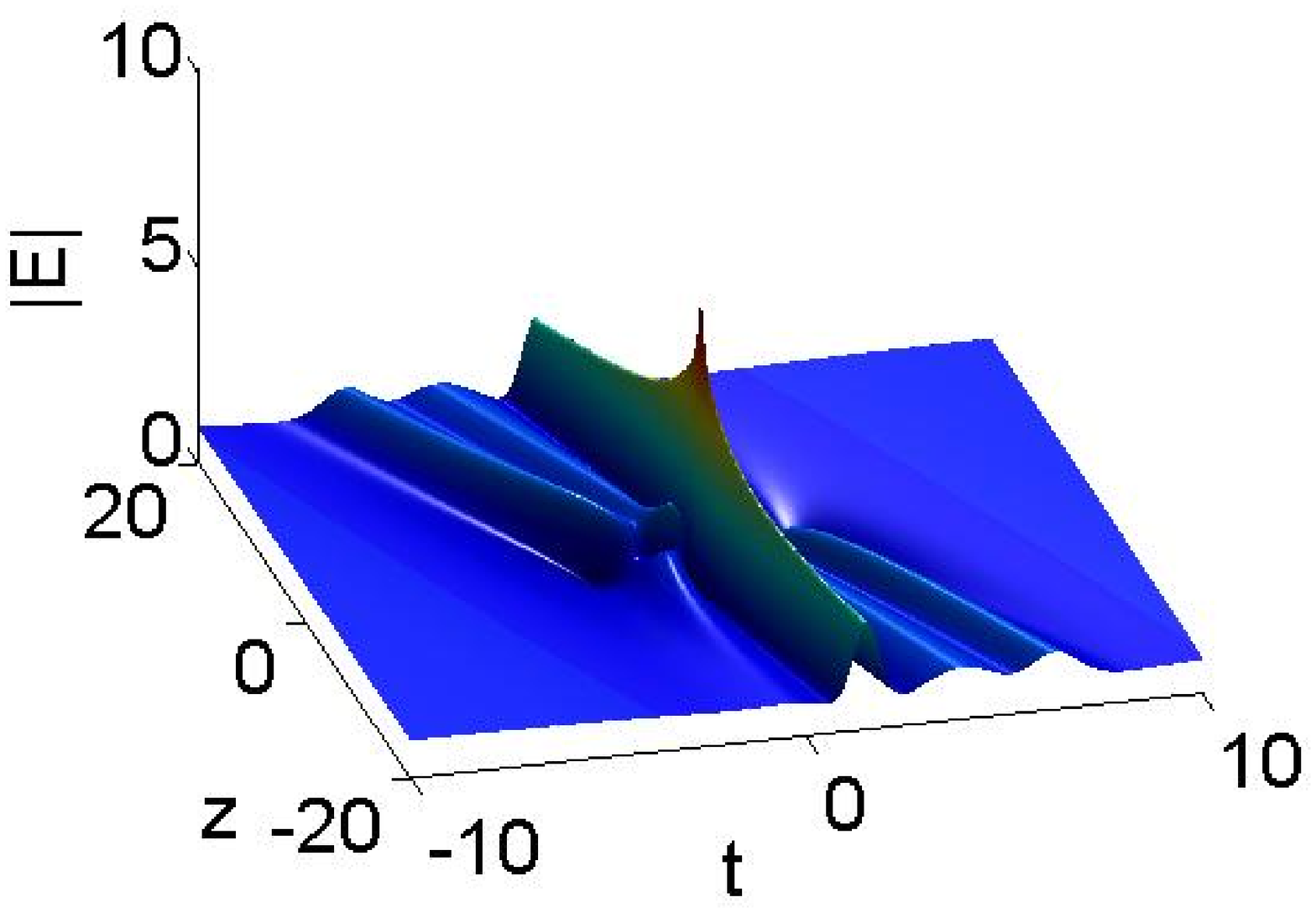}}
\begin{center}
\hskip 1cm $(\rm{a})$ \hskip 6cm $(\rm{b})$
\end{center}
\caption{(a), (b) Density plots of the third-order stationary and nonstationary W-shaped solitons by choosing
$d_{0}=0,b=0$ and $d_{0}=1,b=-2/7$ with $a=1,c=1,\omega=0,s_{1}=0,s_{2}=0$.  }
\label{fig:12}
\end{figure}

For the third-order W-shaped soliton, we omit presenting the lengthy explicit solution with form of the mixture of
higher-order rational polynomials and exponential function. As is shown in Fig. \ref{fig:12}(a),
there are five humps in the $t-z$ distribution plane, viz., $t=0$, $t=\pm1.6562$ and $t=\pm0.5833$,
and their amplitudes are 5, 0.5139 and 1.8991.
Moreover, the six valleys are reached at $t=\pm2.6714$, $t=\pm1.0931$ and $t=\pm0.3238$,
and their amplitudes are all 0. Based on these facts, we can infer that the
$N$th-order stationary W-shaped soliton has $2N-1$ humps and $2N$ valleys.
On the other hand, the third-order nonstationary W-shaped soliton is exhibited in Fig. \ref{fig:12}(b),
the maximum amplitude of it is 7 and arrives at $(0,0)$.

\section{Spectrum analysis and energy of the rogue wave solution}

It is recently recognized that spectrum analysis is a meaningful tool for the prediction and excitation
of the rogue wave \cite{39}. Therefore, in this section, we try to perform the spectrum analysis of the first-order
rogue wave solution. In order to more conveniently reveal the relationship between the spectrum and
state transition condition in comparison with Figs. \ref{fig:1}(a), \ref{fig:2}(a) and \ref{fig:10}(b),
we consider the following rogue wave solution with the special choice of $a=1,c=1,d_{0}=1$ and $\omega=0$ in Eq. (\ref{R1}),
namely,
\begin{equation}\label{E1s}
E[1]_{s}=-\dfrac{100t^2-(16b-24)zt+(32b^2+16b+4)z^2-75+i(112b+32)z}{100t^2-(16b-24)zt+(32b^2+16b+4)z^2+25}e^{i(t+bz)}.
\end{equation}
Then the spectrum of Eq. (\ref{E1s}) yields
\begin{equation}\label{Fouier}
|F(\lambda,z)|=\left|\dfrac{1}{\sqrt{2\pi}}\int_{-\infty}^{\infty}E[1]_{s}(t,z)e^{i\lambda t}dt\right|=\sqrt{2\pi}
\exp\left(-\frac{\sqrt{16(7b+2)^2z^2+625}}{50}|\lambda^{'}| \right),
\end{equation}
where $\lambda^{'}=\lambda-1$.

\begin{figure}[!h]
\centering
\renewcommand{\figurename}{{\bf Fig.}}
{\includegraphics[height=5.5cm,width=5.5cm]{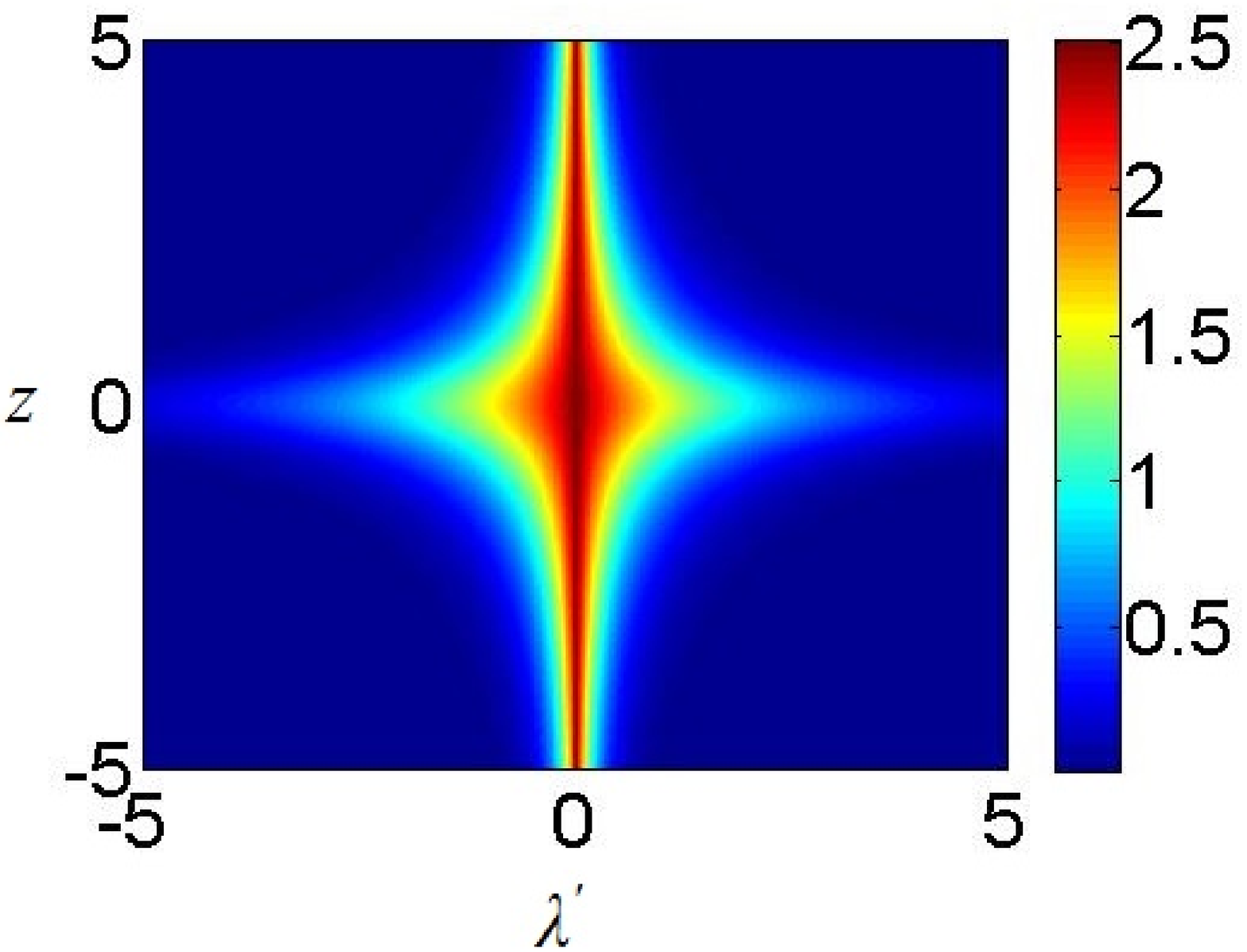}}
{\includegraphics[height=5.5cm,width=5.5cm]{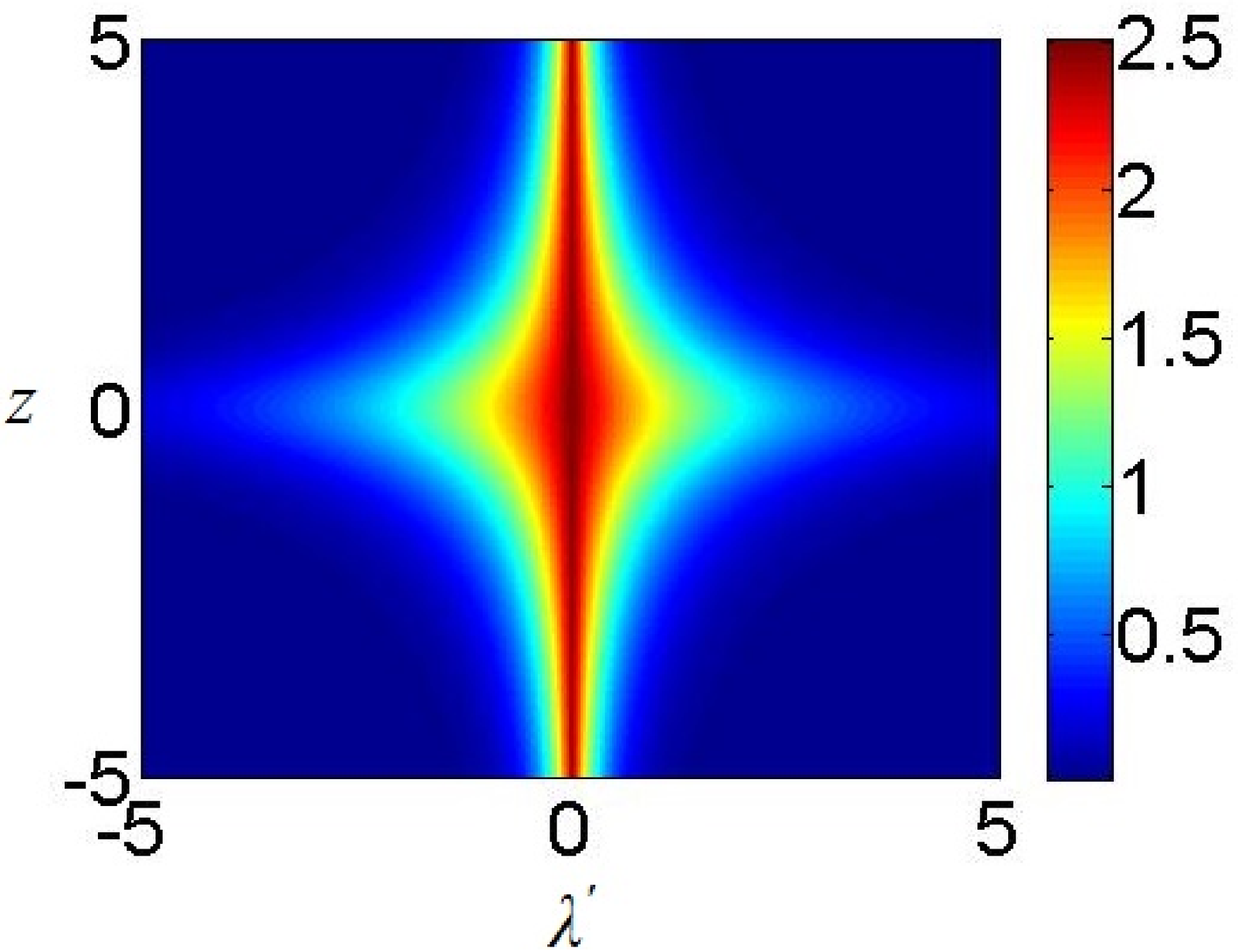}}
{\includegraphics[height=5.5cm,width=5.5cm]{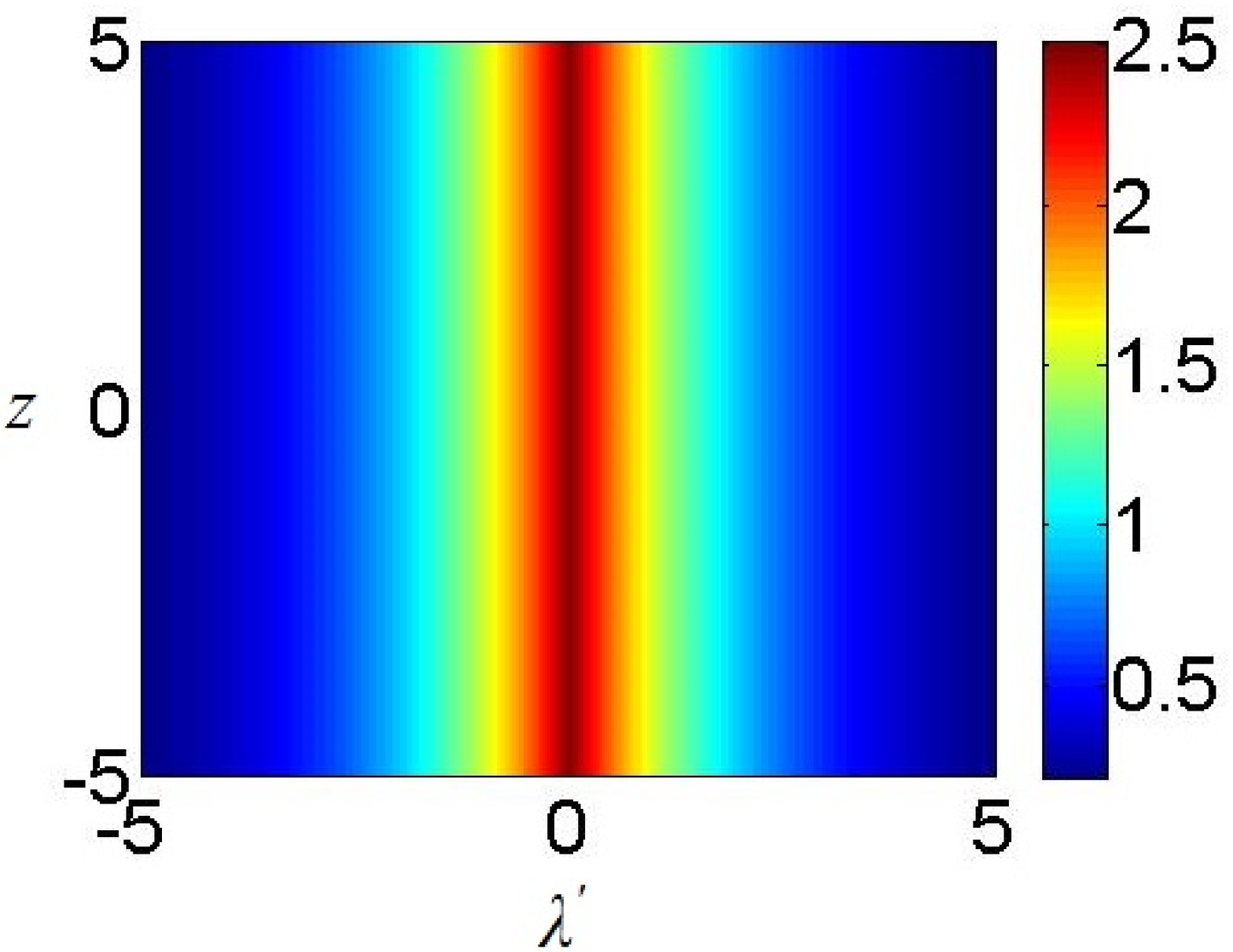}}
\begin{center}
\hskip 1cm $(\rm{a})$ \hskip 5.5cm $(\rm{b})$ \hskip 5.5cm $(\rm{c})$
\end{center}
\caption{Spectrum dynamics $|F(\lambda,z)|$ with (a) $b=3/2$, (b) $b=-3/2$ and (c) $b=-2/7$. }
\label{fig:13}
\end{figure}

Given this definition, one can easily find that when $b\neq-2/7$, the spectrum
is nothing but that of the standard Peregrine soliton which features a triangular shape \cite{39}. We observe that, in
Figs. \ref{fig:13}(a) and \ref{fig:13}(b), the spectrum of the rogue wave solution with $b=-3/2$
gets broadened at the maximally compressed point ($z=0$) in contrast with that
of the rogue wave solution with $b=3/2$. Further,
when taking $b=-2/7$, the spectrum of the nonstationary W-shaped soliton becomes stable and broad,
see Fig. \ref{fig:13}(c).

The evolutional process of the spectrum is completely coincide with the
state transition presented in Figs. \ref{fig:1}(a), \ref{fig:2}(a) and \ref{fig:10}(b). In addition,
we can define the energy of the rogue wave pulse:
\begin{equation}
E_{pulse}(z)=\int_{-\infty}^{+\infty}|E[1]_{s}(t,z)-E[1]_{s}(\pm\infty,z)|^2dt=\frac{100\pi}{\sqrt{16(7b+2)^2z^2+625}}.
\end{equation}
The free parameter $b$ controls the wave width of the energy function. When $b=-2/7$, the rogue wave is converted to
W-shaped soliton, and the energy becomes a constant, i.e. $4\pi$. This fact further verifies
the state transition condition obtained through the MI analysis.

\section{Conclusion and discussions}

In conclusion, by using the generalized DT method, we proposed a compact determinant representation for arbitrary $N$th-order rational solution of the multiple SIT system, which plays an important role in enhancement of the amplification and control of optical waves
compared to the single SIT system.
By virtue of the rational solution, MI and the trace analysis method, we put forward
the optical rogue wave, as well as the stationary and nonstationary
W-shaped solitons structures from first to third order.
We found that the nonstationary W-shaped solitons could only exist in the multiple SIT system instead of the single case.
Further, we demonstrated that the evolutional processes of the spectrum and energy of the rogue wave solution were
completely coincide with the state transition condition given by the MI analysis.

Besides, on the one hand, we note that the $N$th-order Akhmediev breather solution, multi-peak solution and periodic solution of the
multiple SIT system can be obtained by slightly adjusting Eq. (\ref{solu}) with the small complex parameter $f$ being chosen as the nonzero constant. On the other hand, the results presented in this paper can be
directly generalized to the resonant erbium-doped fiber system described by the NLS-multiple SIT equations.
These problems are also interesting and we will summarize the corresponding results
in our recent papers.

\section*{Acknowledgment}
One of the authors X. Wang would like to thank Professor Y. Chen, Doctor L.M. Ling, L.C. Zhao and L. Wang
for their valuable suggestions and discussions.

\section*{Appendix A: Polynomials in Eq. (\ref{15})}
\begin{align}
&F_{2}=1000000t^6-2250000t^4-2812500t^2-384(2b-3)(8b^2+4b+1)^2z^5t+64(8b^2+4b+1)^3z^6~~~~~~\nonumber\\
&~~~+(-256(2b-3)(608b^2+276b+93)t^3+(912384b^3-73728b^2-33408b+16704)t)z^3\nonumber\\
&~~~+(4800(216b^2+52b+61)t^4-36000(32b^2-4b-1)t^2+2412000b^2+2142000b+607500)z^2\nonumber\\
&~~~+(192(216b^2+52b+61)(8b^2+4b+1)t^2-1483776b^4-1625088b^3-764928b^2-160896b\nonumber\\
&~~~-11280)z^4+(-240000(2b-3)t^5+24000(-2b-37)t^3+9000(-142b-27)t)z+703125,\nonumber\\
&G_{2}=-3072(7b+2)(2b-3)(8b^2+4b+1)z^4t+768(7b+2)(8b^2+4b+1)^2z^5+(1536(7b+2)\nonumber\\
&~~~\times(208b^2+76b+43)t^2-2896896b^3-2631168b^2-920448b-82176)z^3+(-76800(2b-3)\nonumber\\
&~~~\times(7b+2)t^3+288000(6b^2+3b+2)t)z^2+(480000(7b+2)t^4-144000(-37b-22)t^2\nonumber\\
&~~~-558000b+252000)z,\nonumber\\
&D_{2}=1000000t^6+750000t^4+1687500t^2-384(2b-3)(8b^2+4b+1)^2z^5t+64(8b^2+4b+1)^3z^6\nonumber\\
&~~~+(-256(2b-3)(608b^2+276b+93)t^3+(3072b^3+792576b^2+625536b+148032)t)z^3\nonumber\\
&~~~+(4800(216b^2+52b+61)t^4+36000(128b^2+84b+21)t^2+684000b^2+558000b+211500)z^2\nonumber\\
&~~~+(192(216b^2+52b+61)(8b^2+4b+1)t^2+48(24b^2-12b-11)^2)z^4+(-240000(2b-3)t^5\nonumber\\
&~~~-24000(42b-23)t^3-9000(-34b-29)t)z+140625.\nonumber
\end{align}

\section*{Appendix B: Polynomials in Eq. (\ref{20})}
\begin{align}
&F_{2}^{'}=1000000t^6-2250000t^4-150000000t^3-2812500t^2+\dfrac{6000000}{16807}d_{0}^5z^5t-337500000t~~~~~~~~~~\nonumber\\
&~~~+\left(\dfrac{15000000}{2401}d_{0}^4t^2+\dfrac{90000}{343}d_{0}^4\right)z^4+\left(\dfrac{20000000}{343}d_{0}^3t^3
-\dfrac{360000}{343}d_{0}^3t-\dfrac{150000000}{343}d_{0}^3\right)z^3\nonumber\\
&~~~+\left(\dfrac{15000000}{49}d_{0}^2t^4-\dfrac{4860000}{49}d_{0}^2t^2-\dfrac{450000000}{49}d_{0}^2t+\dfrac{9427500}{49}d_{0}^2\right)z^2
+\dfrac{1000000}{117649}d_{0}^6z^6\nonumber\\
&~~~+\left(\dfrac{6000000}{7}d_{0}t^5-\dfrac{6120000}{7}d_{0}t^3
-\dfrac{450000000}{7}d_{0}t^2+\dfrac{855000}{7}d_{0}t-\dfrac{553500000}{7}d_{0}\right)z\nonumber\\
&~~~+5625703125,\nonumber
\end{align}

\begin{align}
&G_{2}^{'}=\dfrac{23040000}{49}d_{0}^2z^2t+\dfrac{11520000}{343}d_{0}^3z^3+\left(\dfrac{11520000}{7}d_{0}t^2
+\dfrac{2880000}{7}d_{0}\right)z,\nonumber\\
&D_{2}^{'}=1000000t^6+750000t^4-150000000t^3+1687500t^2+\dfrac{6000000}{16807}d_{0}^5z^5t+112500000t\nonumber\\
&~~~+\left(\dfrac{15000000}{2401}d_{0}^4t^2+\dfrac{3630000}{2401}d_{0}^4\right)z^4+\left(\dfrac{20000000}{343}d_{0}^3t^3
+\dfrac{11640000}{343}d_{0}^3t
-\dfrac{150000000}{343}d_{0}^3\right)z^3\nonumber\\
&~~~+\left(\dfrac{15000000}{49}d_{0}^2t^4+\dfrac{13140000}{49}d_{0}^2t^2
-\dfrac{450000000}{49}d_{0}^2t+\dfrac{5287500}{49}d_{0}^2\right)z^2+\dfrac{1000000}{117649}d_{0}^6z^6\nonumber\\
&~~~+\left(\dfrac{6000000}{7}d_{0}t^5+840000d_{0}t^3-\dfrac{450000000}{7}d_{0}t^2+\dfrac{1215000}{7}d_{0}t-\dfrac{103500000}{7}d_{0}
\right)z\nonumber\\
&~~~+5625140625.\nonumber
\end{align}

\end{document}